\documentstyle[aps,prd,epsfig]{revtex}

\newcommand{\NPA}[3]{Nucl.\ Phys.\ {\bf A#1},\ #2 (#3)}
\newcommand{\NPB}[3]{Nucl.\ Phys.\ {\bf B#1},\ #2 (#3)}

\newcommand{\PLB}[3]{Phys.\ Lett.\ B\ {\bf #1},\ #2 (#3)}

\newcommand{\PRL}[3]{Phys.\ Rev.\ Lett.\ {\bf #1},\ #2 (#3)}

\newcommand{\PRD}[3]{Phys.\ Rev.\ D\ {\bf #1},\ #2 (#3)}

\newcommand{\ibid}[3]{{\bf #1},\ #2 (#3)}

\renewcommand\a{\alpha}
\renewcommand\b{\beta}
\newcommand\g{\gamma}
\renewcommand\d{\delta}
\newcommand\e{\epsilon}

\renewcommand\k{\kappa}

\newcommand\m{\mu}
\newcommand\n{\nu}
\newcommand\x{\xi}
\newcommand\p{\pi}

\newcommand\D{\Delta}

\newcommand{\non}{\nonumber\\}

\newcommand{\be}{\begin{equation}}
\newcommand{\ee}{\end{equation}}
\newcommand{\bea}{\begin{eqnarray}}
\newcommand{\eea}{\end{eqnarray}}
\newcommand{\ba}[1]{\begin{array}{#1}}
\newcommand{\ea}{\end{array}}

\newcommand{\eqrf}[1]{Eq.\ (\ref{#1})}

\newcommand{\bm}[1]{\mbox{\boldmath${#1}$}}
\newcommand{\uq}{\hat{\mathbf{q}}} 
\newcommand{\uk}{\hat{\mathbf{k}}}

\newcommand{\vg}{\bm{\gamma}}
\newcommand{\gperp}{\bm{\gamma}_{\perp}}
\newcommand{\vf}{\bm{\phi}}
\newcommand{\vphi}{\bm{\varphi}}

\begin{document}

\title{When the Transition Temperature in Color Superconductors 
is Not Like in BCS Theory}

\author{Andreas Schmitt}
\address{Institut f\"ur Theoretische Physik, 
J.W. Goethe-Universit\"at D-60054 Frankfurt/Main, Germany \\
E-mail: aschmitt@th.physik.uni-frankfurt.de}

\author{Qun Wang}
\address{Institut f\"ur Theoretische Physik, 
J.W. Goethe-Universit\"at, D-60054 Frankfurt/Main, Germany, \\
and Physics Department, Shandong University, Jinan, Shandong, 250100, 
P.R. China \\
E-mail: qwang@th.physik.uni-frankfurt.de}

\author{Dirk H. Rischke}

\address{Institut f\"ur Theoretische Physik, 
J.W. Goethe-Universit\"at, D-60054 Frankfurt/Main, Germany \\
E-mail: drischke@th.physik.uni-frankfurt.de}

\date{\today}
\maketitle

\begin{abstract}

We study color superconductivity with $N_f=1,2,$ and 3 massless flavors
of quarks. We present a general formalism to derive and solve the gap 
equations for condensation in the even-parity channel.
This formalism shows that the leading-order contribution
to the gap equation is unique for all color superconductors studied
here, and that differences arise solely at the subleading order.
We discuss a simple method to compute subleading contributions from the
integration over gluon momenta in the gap equation. Subleading
contributions enter the prefactor of the color-superconducting
gap parameter.
In the case of color-flavor and color-spin locking we identify
further corrections to this prefactor arising from the two-gap structure
of the quasiparticle excitations.
Computing the transition temperature, $T_c$, where the 
color-superconducting condensate melts, we find that these 
contributions lead to deviations from
the BCS behavior $T_c\simeq 0.57\,\phi_0$, where $\phi_0$ is the 
magnitude of the zero-temperature gap at the Fermi surface.  
\end{abstract}

\maketitle

\section{Introduction and Conclusions}

Cold and dense quark matter is a color superconductor
\cite{bailin,alford1}.
At asymptotically large quark density or, equivalently,
quark chemical potential $\mu$, asymptotic freedom \cite{asymp}
implies that the strong coupling constant $g$ becomes
small. In this case one can reliably compute the
color-superconducting gap parameter to leading and subleading order
in $g$ from a gap equation derived within the framework of QCD 
\cite{bailin,QCDgapeq,rischke1}. 
For instance, in a color superconductor with 
$N_f = 2$ massless flavors of quarks 
(commonly called the ``2SC'' phase), 
the value of the gap at the Fermi surface and at zero temperature is
\be \label{phi02SC}
\phi_0^{\rm 2SC}=2 \, \tilde{b} \, b_0'\, \mu\, \exp\left(
-\frac{\pi}{2\, \bar{g}}\right) \,\, ,
\ee
where 
\be \label{constants}
\bar{g} \equiv \frac{g}{3 \sqrt{2}\, \pi} \,\, , \qquad
\tilde{b} \equiv 256 \pi^4 \left(\frac{2}{N_f g^2} \right)^{5/2}\,\, , \qquad
b_0' \equiv \exp\left(-\frac{\pi^2+4}{8}\right)
\,\, .
\ee
The term in the exponent of Eq.\ (\ref{phi02SC}) was
first computed by Son \cite{son}. It arises 
from the exchange of almost static magnetic gluons.
The factor $\tilde{b}$ in front of the exponential originates from 
the exchange of static electric and non-static magnetic gluons
\cite{QCDgapeq,rischke1}.
The prefactor $b_0'$ is due to the quark self-energy \cite{ren,wang}.

In color superconductors with $N_f=1$ and 3 flavors, various other
prefactors may arise \cite{schaefer2,schaefer}, but the exponential
$\exp[-\pi/(2\bar{g})]$ remains the same. 
As will be demonstrated in this paper, this is not an accident, but
due to the fact that the leading-order contribution to the
QCD gap equation does not depend on the 
detailed color, flavor, and Dirac structure of the 
color-superconducting order parameter. This structure
only enters at subleading order, and we provide
a simple method to extract these subleading contributions.

Let us briefly recall what the terms
``leading,'' ``subleading,'' and ``sub-subleading order'' mean
in the context of the QCD gap equation \cite{wang}.
Due to the non-analytic dependence of $\phi_0$ on 
the strong coupling constant $g$ 
one cannot apply the naive perturbative counting scheme
in powers of $g$ in order to identify
contributions of different order.
In the QCD gap equation there are also logarithms of the form
$\ln (\m/\phi_0)$, which are $\sim 1/g$ due to
Eq.\ (\ref{phi02SC}) and thus may cancel simple powers of $g$.
A detailed discussion of the resulting, modified power-counting scheme was 
given in the introduction of Ref.\ \cite{wang} and need not be
repeated here. In short,
leading-order contributions in the QCD gap equation are
due to the exchange of almost static magnetic gluons and
are proportional to $g^2\, \phi_0\, \ln^2(\m/\phi_0) \sim \phi_0$.
They determine the argument of the exponential
in Eq.\ (\ref{phi02SC}).
Subleading-order contributions are due to the exchange of
static electric and non-static magnetic gluons and are 
$\sim g^2 \, \phi_0 \, \ln (\m /\phi_0) \sim g \, \phi_0$.
They determine the prefactor of the exponential in Eq.\ (\ref{phi02SC}).
Finally, sub-subleading contributions 
arise from a variety of sources and, at present, cannot be systematically
calculated. They are proportional to $g^2\, \phi_0$ and
constitute $O(g)$ corrections to the prefactor in Eq.\ (\ref{phi02SC}).
It was argued that also gauge-dependent terms
enter at this order \cite{gauge}. This is, of course,
an artefact of the mean-field approximation which was used to derive
the QCD gap equation \cite{rischke2}. On the quasiparticle mass shell,
the true gap parameter is in principle a physical observable and
thus cannot be gauge dependent.

In color superconductors, the mass shell of a quasiparticle
is determined by its excitation energy
\be \label{excite}
\e_{k,r}(\phi)=\left[(k-\mu)^2+\lambda_r \,|\phi(\e_{k,r},k)|^2
\right]^{1/2} \,\, ,
\ee
where $k\equiv |{\bf k}|$ is the modulus of the 3-momentum of the
quasiparticle, and $\phi(\e_{k,r},k)$ is
the gap function on the quasiparticle mass shell. 
The index $r$ labels possible excitation branches in the superconductor,
which differ by the value of the constant $\lambda_r$.
At the Fermi surface, $k=\mu$, the excitation 
of a quasiparticle -- quasiparticle-hole pair costs an energy
$2\,\e_{\mu,r}(\phi)=2\,\sqrt{\lambda_r}\,\phi_0$.
The true energy gap is therefore $\sqrt{\lambda_r}\,\phi_0$. 

At first sight the introduction of the constant $\lambda_r$ appears
somewhat awkward. The advantage is that it allows to generalize
Eq.\ (\ref{excite}) to different color-superconducting systems.
For example, in a two-flavor color superconductor, 
quarks of two colors form Cooper 
pairs with total spin zero, while the third color
remains unpaired \cite{bailin}. Consequently, there are two different 
excitation energies, $\e_{k,1}$ and $\e_{k,2}$. Four quasiparticle
excitations have $\lambda_1=1$, with gap $\phi_0$,  
while two have $\lambda_2=0$, corresponding to the unpaired quarks. 
These are so-called ``ungapped'' excitations. At the Fermi surface,
it costs no energy to excite them. 
In a three-flavor color superconductor,
with color-flavor locking (CFL) \cite{alford2}, all nine quark 
colors and flavors
form Cooper pairs, but there are still two distinct branches
of fermionic excitations. The first, with $\lambda_1=4$, occurs
with degeneracy one, while the other, with $\lambda_2=1$, has
degeneracy eight. The gap corresponding to the first excitation
has magnitude $2\phi_0$, while for the other eight the size of 
the gap is $\phi_0$. A similar two-gap structure also appears 
in the color-spin locked (CSL) phase of a one-flavor 
color superconductor \cite{schaefer}.
However, here the first excitation, with $\lambda_1=4$, has a four-fold 
degeneracy, while  the second, with $\lambda_2=1$, has an  
eight-fold degeneracy. 

In this paper we aim to clarify the similarities and differences
between various color-superconducting systems. To this end 
we systematically study six different cases.
The first two cases are spin-zero color superconductors with
(i) two flavors of massless quarks 
and (ii) three flavors of massless quarks  
in the CFL phase. The other four cases deal with one massless quark 
flavor. In this case, the condensate has spin one \cite{iwasaki}.
Similar to Helium-3 \cite{vollhardt}, this allows for a multitude of 
different phases,
distinguished by the symmetries of the order parameter \cite{schaefer}.
We only focus on (iii) the CSL phase with longitudinal and transverse
gaps, (iv) the CSL phase with longitudinal gap only, (v) 
the CSL phase with transverse gap only, and (vi) the
polar phase. In this context, ``longitudinal'' 
and ``transverse'' refers to pairing of quarks with the same or
different chiralities, respectively \cite{rischke1}. The cases
(iv) and (v) can be considered separately because, as we shall
show below, longitudinal and transverse gaps do not induce each other.
In all six cases we only consider condensation in even-parity channels,
because these are favored by effects which explicitly break
the $U(1)_A$ symmetry of the QCD Lagrangian.

We show that in all six cases the gap equation has the 
general form
\begin{equation}
\phi(\e_{k,r},k)=\bar{g}^2
\int_0^\delta d(q-\mu)\sum_s a_s \; Z(\e_{q,s})\,
\frac{\phi(\e_{q,s},q)}{\e_{q,s}}\;
\tanh\left(\frac{\e_{q,s}}{2T}\right)\;\frac{1}{2}\ln\left(\frac
{b^2\mu^2}{|\e_{q,s}^2-\e_{k,r}^2|}\right) \,\, .
\label{gap1}
\end{equation}
The sum over $s$ runs over all distinct branches of
fermionic excitations with energy $\e_{q,s}$ in the color
superconductor. For the systems considered here, there
are only two such branches, such that $s=1$ or 2.
The coefficients $a_s$ are positive numbers, obeying
the constraint 
\be \label{constraint}
\sum_{s=1}^2 a_s=1\,\, .
\ee 
In the first four columns of Table \ref{table1} 
we display the values of $\lambda_s$ and $a_s$ for the six
cases studied here.
There are ungapped excitations ($\lambda_s=0$) in the cases (i),
(iv), (v), and (vi), while
all excitations are gapped in the cases (ii) and (iii). 
For ungapped excitations, the corresponding $a_s$ vanishes, and thus these 
do not appear in the gap equation. This is natural, because ungapped
excitations should not affect the value of the color-superconducting
gap.
The constants are identical for the cases (ii) and (iii), 
$\lambda_1 = 4$ and $\lambda_2 = 1$.
The coefficients $a_s$ also assume the same values, $1/3$ and $2/3$,
but the association of these values 
with the corresponding constants $\lambda_s$ is reversed
in case (iii) as compared to case (ii).

The occurrence of the wave function renormalization factor $Z(\e_{q,s})$ 
in a gap equation of the type (\ref{gap1}) was first discussed 
in Ref.\ \cite{wang}.
The constant $b$ in Eq.\ (\ref{gap1}) is defined as
\begin{equation}
\label{b} 
b\equiv \tilde{b}\,\exp(-d) \,\, ,
\end{equation}
with $\tilde{b}$ from Eq.\ (\ref{constants}), and $d$ a constant
of order one.
The constant $d$ originates from subleading contributions to the gap equation.
For spin-zero condensates, $d=0$, due to an accidental 
cancellation of some of the subleading terms arising
from static electric and non-static magnetic gluon exchange.
In the spin-one cases, this cancellation does not occur 
and, consequently, $d\neq 0$. 

In this paper we present a simple method to extract the value
of the constant $d$ without actually solving a gap equation.
This method utilizes the fact that, to subleading order,
the integration over gluon
momenta in the QCD gap equation can be written as a sum
of a few integrals multiplied by constants. Only these
constants depend on the detailed color, flavor, and Dirac structure
of the order parameter. The integrals are generic for all
cases studied here and have to be computed only once.
The precise numerical values for $d$ are listed in the fifth column of Table
\ref{table1}. 

The fact that we can write the gap equation in all six cases
in the form (\ref{gap1}) is nontrivial. It means that the 
leading contribution to the gap equation is {\em unique}. 
If it were not, then the prefactor
of the gap integral would be different for each case.
In other words, the contribution of almost static magnetic gluons
to the gap equation is universal in the sense that it is
independent of the detailed color, flavor, and Dirac structure
of the color-superconducting order parameter.
Differences between the six cases studied here occur
at subleading order. Only at this order the specific
structure of the order parameter 
is important and leads to different values
for the constant $d$ in Eq.\ (\ref{b}).

We solve the gap equation (\ref{gap1}) at zero temperature and
compute the value of the gap function at the Fermi surface, $\phi_0$.
In all cases studied here, we can write the result in the form
\be \label{phi0}
\phi_0 = 2\,\, b \, b_0' \, \m \, \exp\left(- \frac{\pi}{2 \, \bar{g}}
\right)\, \left( \lambda_1^{a_1} \, \lambda_2^{a_2} \right)^{-1/2}\,\, .
\ee
(Remember that $0^0 \equiv 1$.) From 
this equation and Eq.\ (\ref{phi02SC}) one 
immediately determines $\phi_0$ in units
of the gap in the 2SC phase,
\be \label{ratio}
\frac{\phi_0}{\phi_0^{\rm 2SC}} = \exp(-d) \, 
\left( \lambda_1^{a_1} \, \lambda_2^{a_2} \right)^{-1/2}\,\, .
\ee 
This ratio is given in the sixth column of Table \ref{table1}.
For spin-one color superconductors, $d$ is positive, 
and the exponential factor leads to a tremendous
suppression of the gap by factors $e^{-4.5} \simeq 10^{-2}$
to $e^{-6} \simeq 2.5\times 10^{-3}$ relative to the 
spin-zero gap \cite{ren,schaefer}. 
In contrast to the value of $d$, the additional factor 
$\left( \lambda_1^{a_1} \lambda_2^{a_2} \right)^{-1/2}$
in Eq.\ (\ref{ratio}) cannot be simply read off from the
subleading contributions in the gap equation, but 
only follows from the explicit solution.
It is different from 1 for color superconductors 
with two distinct
branches of {\em gapped\/} quasiparticle excitations.
In this case, this factor
further reduces $\phi_0$ as compared to the 2SC case. 

The factor $\left( \lambda_1^{a_1} \lambda_2^{a_2} \right)^{-1/2}$
is also different from 1 in case (v) where
$\lambda_1 = 2$. However, according to Eq.\
(\ref{excite}) the ``true'' gap is $\sqrt{\lambda_1}\, \phi_0= \sqrt{2}\,
\phi_0$, and not $\phi_0$. Therefore, the ratio of the true gap
to the gap in the 2SC case is just $\exp(-d)$. 
In order to indicate this, in Table \ref{table1}
we put the factor $2^{-1/2}$ arising from 
Eq.\ (\ref{ratio}) in parentheses.

\begin{table} \label{table1} 
\begin{tabular}[t]{|cc||c|c|c|c|c|c|c|} 
&& $\lambda_1$ & $\lambda_2$ & $a_1$ & $a_2$ & $d$ & 
$\phi_0/\phi_0^{\rm 2SC}$ & $T_c/(e^\gamma\phi_0/\pi)$ \\ 
\hline\hline 
(i) & 2SC & 1 & 0 & 1 & 0& 0& 1& 1\\
\hline
(ii) & CFL & 4 & 1 & $1/3$ & $2/3$ & 0
& $2^{-1/3}$ & $2^{1/3}$ \\
\hline
(iii) & CSL (transv.+long.) & 4 & 1 & $2/3$ & $1/3$
& 5& $2^{-2/3}e^{-d}$ & $2^{2/3}$\\
\hline
(iv) & CSL (long.) & 1 & 0 & 1 & 0 & 6& $e^{-d}$& 1\\
\hline
(v) & CSL (transv.) & 2 & 0 & 1 & 0 & 9/2& $(2^{-1/2}) \, e^{-d}$& 
$(2^{1/2})\, 1$\\
\hline
(vi) & polar & 1 & 0 & 1 & 0 & $\,\frac{3}{2}(3+\cos^2\vartheta)\,$
& $e^{-d}$& 1
\end{tabular}
\caption{The constants $\lambda_s$, $a_s$, $d$, the ratio 
$\phi_0/\phi_0^{\rm 2SC}$, and the ratio $T_c/\phi_0$ normalized 
to its BCS value. In case (vi), $\vartheta$ is the angle between the
direction of the spin-one condensate and the 3-momentum of the 
quarks in the Cooper pair, see Sec.\ \ref{polarphase}. In case (v), 
the factors in parentheses do not occur if
$\phi_0$ is replaced by the true gap $\sqrt{\lambda_1}\, \phi_0$.}
\end{table}

Finally, we discuss the transition temperature $T_c$, where the 
color-superconducting condensate melts. 
We find
\be
\label{Tc}
\frac{T_c}{\phi_0}=\frac{e^\gamma}{\pi} \, 
\left( \lambda_1^{a_1} \, \lambda_2^{a_2} \right)^{1/2} 
\simeq 0.57 \, \left( \lambda_1^{a_1} \, \lambda_2^{a_2} \right)^{1/2}\,\, ,
\ee
where $\gamma\simeq 0.577$ is the Euler-Mascheroni constant.
In the cases (i), (iv), and (vi), where there is only one
gapped quasiparticle excitation,
$( \lambda_1^{a_1} \, \lambda_2^{a_2})^{1/2} = 1$,
and we recover the relation $T_c/\phi_0 \simeq 0.57$
well known from BCS theory \cite{bcs}.
Its validity for QCD with $N_f =2$ flavors of massless
quarks was first demonstrated in Refs.\ \cite{rischke1,wang}.
In case (v), $( \lambda_1^{a_1} \, \lambda_2^{a_2})^{1/2} = \sqrt{2}$,
but this factor is absent if we rescale $\phi_0$ in Eq.\ (\ref{Tc})
by $\sqrt{\lambda_1}$ to obtain the true gap.
Therefore, also in this case the BCS relation between the
zero-temperature gap and the critical temperature is valid.
In the cases (ii) and (iii) there are two distinct gapped quasiparticle
excitations, and consequently two gaps, $\sqrt{\lambda_1} \, \phi_0 =
2 \, \phi_0$
and $\sqrt{\lambda_2} \,\phi_0 = \phi_0$. 
The BCS relation $T_c/\phi_0 = e^\gamma/\pi$ is violated by the 
additional factor
$( \lambda_1^{a_1} \, \lambda_2^{a_2} )^{1/2} > 1$.

In order to elucidate the deviations from the BCS relation,
in the last column of Table
\ref{table1} we present our results for $T_c$
in units of the critical temperature expected
from BCS theory.
Apparently, the two-gap structure in the cases (ii) and (iii)
is responsible for the observed deviations.
It would be interesting to observe similar behavior
in other weak-coupling superconductors with
more than one gapped excitation branch.
Note that, although $T_c/\phi_0$ is different than in BCS theory,
the absolute values of $T_c$ do not change. If the
energy scale is set by $\phi_0^{\rm 2SC}$, then
$T_c/ \phi_0^{\rm 2SC} = \exp(-d)$, because the factor
$(\lambda_1^{a_1}\, \lambda_2^{a_2})^{-1/2}$
in Eq.\ (\ref{ratio}) simply cancels the factor
$(\lambda_1^{a_1}\, \lambda_2^{a_2})^{1/2}$ in Eq.\ (\ref{Tc}).

The remainder of this paper is organized as follows.
In Section \ref{2} we show that the gap equations for all six
cases considered in this paper is of the form (\ref{gap1}). 
We explain the origin of the constants $\lambda_s$ as 
eigenvalues of an operator constructed from the 
color-superconducting gap matrix.
We also present a simple method to compute subleading corrections
to the gap arising from the integration over gluon momenta
in the gap equation, leading to the suppression factor $\exp(-d)$
in Eq.\ (\ref{b}). In Section \ref{solution} we solve the gap equation
(\ref{gap1}) at zero temperature and explain the occurrence of the 
additional factor $(\lambda_1^{a_1} \, \lambda_2^{a_2})^{-1/2}$
in Eq.\ (\ref{phi0}).
Finally, in Section \ref{transtemp} we compute the critical 
temperatures $T_c$.

Our convention for the metric tensor is 
$g^{\mu\nu}=\mbox{diag}\{1,-1,-1,-1\}$. 
Our units are $\hbar=c=k_B=1$. Four-vectors
are denoted by capital letters, 
$K\equiv K^\mu=(k_0,{\bf k})$, 
and $k\equiv|{\bf k}|$, while $\uk\equiv{\bf k}/k$.
We work in the imaginary-time formalism, i.e., $T/V \sum_K \equiv
T \sum_n \int d^3{\bf k}/(2\pi)^3$, where $n$ labels the Matsubara 
frequencies $\omega_n \equiv i k_0$. For bosons, $\omega_n=2n \pi T$,
for fermions, $\omega_n=(2n+1) \pi T$.

\section{Gap equations}\label{2}

\subsection{General derivation} \label{generalder}

In fermionic systems at non-zero density, 
it is advantageous to treat fermions and 
charge-conjugate fermions as independent degrees of
freedom and to work in the so-called Nambu-Gorkov basis. 
In this basis, the full inverse fermion 
propagator is defined as 
\be
S^{-1}\equiv\left(\ba{cc} S_{11}^{-1}&S_{12}^{-1}\\
                S_{21}^{-1}&S_{22}^{-1} 
        \ea \right)
=\left(\ba{cc} {S^0}_{11}^{-1} + \Sigma_{11} & \Sigma_{12}\\ 
               \Sigma_{21} & {S^0}_{22}^{-1} + \Sigma_{22} 
       \ea \right)\;,
\label{S1-2}
\ee
where ${S^0}_{11}$ is the propagator for free fermions, 
${S^0}_{22}$ the propagator for free 
charge-conjugate fermions. In momentum space 
and for massless quarks,
\be 
{S^0}_{11}(K)=\left( \g^{\m}K_{\m} + \m \g_0 \right)^{-1}\;,\;\;
{S^0}_{22}(K)=\left( \g^{\m}K_{\m} - \m \g_0 \right)^{-1}\;, 
\ee
where $\g ^{\m}$ are the Dirac matrices. 
The $11$ component of the self-energy, 
$\Sigma_{11}$, is the standard one-loop 
self-energy for fermions; similarly, $\Sigma_{22}$ 
is the self-energy for charge-conjugate fermions. 
In Ref.\ \cite{wang} it was shown that, in order to solve
the gap equation to subleading order, it is 
permissible to approximate these self-energies by 
\be \label{selfenergy}
\Sigma(K)\equiv \Sigma_{11}(K)=\Sigma_{22}(K)\simeq \gamma_0\, \bar{g}^2\,k_0\,
\ln\frac{M^2}{k_0^2}\,\, ,
\ee
where $M^2=(3\pi/4)m_g^2$; the zero-temperature gluon mass
parameter (squared) is $m_g^2=N_f g^2 \mu^2/(6\pi^2)$. 
The $21$ component of the self-energy, 
$\Sigma_{21}$, which was denoted $\Phi^+$ in 
\cite{rischke1}, is the gap matrix in a superconductor, 
while $\Sigma_{12}=\g _0\Sigma_{21}^{\dagger}\g _0$. 

Inverting \eqrf{S1-2} one obtains the full 
fermion propagator $S$. The 11 component,
\be
S_{11}=({S^0}_{22}^{-1} + \Sigma_{22})
\left[ ({S^0}_{11}^{-1} + \Sigma_{11}) 
({S^0}_{22}^{-1} + \Sigma_{22})
-{\cal C}\,\right]^{-1}\; ,
\label{S11}
\ee
is the full quasiparticle propagator, where we defined
\be
\label{C}
{\cal C}\equiv\Sigma_{12} ({S^0}_{22}^{-1} + \Sigma_{22})^{-1}
\Sigma_{21}({S^0}_{22}^{-1} + \Sigma_{22}) \,\, .
\ee
The 21 component is the so-called ``anomalous'' propagator. It is given by 
\be \label{S21}
S_{21}=-({S^0}_{22}^{-1}+\Sigma_{22})^{-1} \, \Sigma_{21} \, S_{11} \,\, . 
\ee
In all cases considered here, the gap matrix can be written as
\be \label{gm2SC}
\Sigma_{21}(K)=\sum_{e=\pm}  \phi^e(K)\, {\cal M}_{\bf k} \, \Lambda^e_{\bf k} 
\,\, ,
\ee
where $\phi^e(K)$ is the gap function, ${\cal M}_{\bf k}$ is a matrix
defined by the symmetries of the color-superconducting condensate, and 
$\Lambda_{\bf k}^e=(1+e\gamma_0 \vg \cdot \uk)/2$, 
$e=\pm$, are projectors onto states of positive or negative energy. 
In general, ${\cal M}_{\bf k}$ is a matrix in color, flavor, and Dirac
space, and is constructed such that 
\be \label{M}
[{\cal M}_{\bf k},\Lambda^e_{\bf k}] = 0 \,\, .
\ee
With the gap matrix (\ref{gm2SC}), the operator ${\cal C}(K)$ assumes
the form
\be \label{C2}
{\cal C}(K)=\sum_e |\phi^e(K)|^2 \, L_{\bf k}\, \Lambda^{-e}_{\bf k} \,\, ,
\ee
where
\be \label{Ldef}
L_{\bf k}\equiv\gamma_0\, {\cal M}^\dagger_{\bf k}{\cal M}_{\bf k}\, 
\gamma_0\,\, .
\ee
Note that also $[L_{\bf k}, \Lambda_{\bf k}^e] =0$.
Since $L_{\bf k}$ is hermitian, it has 
real eigenvalues and can be expanded in terms of a complete set of
orthogonal projectors ${\cal P}_{\bf k}^r$,
\be \label{L}
L_{\bf k} = \sum_r \lambda_r \, {\cal P}_{\bf k}^r \,\, ,
\ee 
where $\lambda_r$ are the eigenvalues of $L_{\bf k}$. Our choice
of the symbol $\lambda_r$ is judicial: it will turn out that they
are identical with the constants $\lambda_r$ appearing in the
quasiparticle excitation energy (\ref{excite}) and which are listed 
in Table \ref{table1}. 
In Appendix \ref{AppA} we determine the
eigenvalues of $L_{\bf k}$ and their degeneracy for the six 
color-superconducting systems studied here.

In all cases considered in this paper, 
there are only two distinct
eigenvalues, so that one can easily 
express the two corresponding projectors in terms of 
$L_{\bf k}$,
\be \label{proj}
{\cal P}_{\bf k}^{1,2}=\frac{L_{\bf k}
-\lambda_{2,1}}{\lambda_{1,2}-\lambda_{2,1}} \,\, .
\ee     
Obviously, these projectors also commute with the energy projectors, 
$[{\cal P}_{\bf k}^{1,2}, \Lambda_{\bf k}^e] =0$.
 
The next step is to compute the full quasiparticle propagator $S_{11}$.
The inversion of the term in brackets in Eq.\ (\ref{S11}) is 
particularly simple, because the four projectors ${\cal P}_{\bf k}^{1,2}
\Lambda_{\bf k}^\pm$ are orthogonal and form a complete set in color, 
flavor, and Dirac space. With Eqs.\ (\ref{selfenergy}), (\ref{S11}), 
(\ref{C2}), and (\ref{L})
we obtain
\be \label{fullprop} 
S_{11}(K)=\left[{S_{22}^0}^{-1}(K)+\Sigma_{22}(K)\right]\sum_{e,r}
{\cal P}_{\bf k}^r\, \Lambda_{\bf k}^{-e}\, \frac{1}{\left[k_0/Z(k_0)\right]^2
-\left[\e_{k,r}^e(\phi^e)\right]^2} \,\, ,
\ee
where
\be
Z(k_0)\equiv\left(1+\bar{g}^2\,\ln\frac{M^2}{k_0^2}\right)^{-1}
\ee
is the wave function renormalization factor introduced in Ref.\ 
\cite{manuel} and 
\be \label{excite2}
\e_{k,r}^e(\phi^e)\equiv \left[(k-e\mu)^2+\lambda_r|\phi^e|^2\right]^{1/2}
\ee
are the excitation energies for quasiparticles, $e=+$, see Eq.\ 
(\ref{excite}), or quasi-antiparticles, $e=-$.

In order to compute the anomalous propagator $S_{21}$, we insert 
Eq.\ (\ref{fullprop}) into Eq.\ (\ref{S21}) and employ
Eq.\ (\ref{gm2SC}). The result is
\be \label{S212SC}
S_{21}(K)=-\sum_{e,r} \gamma_0 \, {\cal M}_{\bf k}\, 
\gamma_0\, {\cal P}_{\bf k}^r \Lambda_{\bf k}^{-e}
\, \frac{\phi^e(K)}{\left[k_0/Z(k_0)\right]^2-
\left[\e_{k,r}^e(\phi^e)\right]^2} \,\, . 
\ee

In the mean-field approximation \cite{rischke2}, $\Sigma_{21}$ obeys the gap
equation \cite{wang}
\be \label{gap2}
\Sigma_{21}(K)=g^2\,\frac{T}{V}\sum_Q
\D_{\m\n}^{ab}(K-Q) \, \g^{\m}T_a^T \,
S_{21}(Q)\,\g^{\n}T_b \,\, ,
\ee
where $T_a$ are the 
Gell-Mann matrices (times a factor 1/2) and $\D_{\m\n}^{ab}$
is the gluon propagator.

To derive the gap equation for the gap function $\phi^e(K)$, we 
insert Eq.\ (\ref{S212SC}) into Eq.\ (\ref{gap2}), 
multiply both sides from the right
with ${\cal M}^\dagger_{\bf k}\, \Lambda_{\bf k}^e$
and trace over color, flavor, and Dirac space. 
To subleading order in the gap equation, it is permissible to use the 
gluon propagator in the Hard-Dense-Loop (HDL) approximation 
\cite{rischke6}, where it is diagonal in adjoint color space, 
$\Delta^{\m\n}_{ab}=\delta_{ab}\, \Delta^{\m\n}$. We obtain
\be \label{gap3}
\phi^e(K)=g^2\frac{T}{V}\sum_{Q}\sum_{e',s} 
\frac{\phi^{e'}(Q)}{\left[q_0/Z(q_0)\right]^2-
\left[\e_{q,s}^{e'}(\phi^{e'})\right]^2} \, \Delta^{\m\n}(K-Q) \,
{\cal T}_{\m\n}^{ee',s}({\bf k},{\bf q}) \,\, ,
\ee 
where 
\be \label{T2SC}
{\cal T}_{\m\n}^{ee',s}({\bf k},{\bf q})=-\frac{ {\rm Tr}
\left[\gamma_\m \, T_a^T \, \gamma_0 \,{\cal M}_{\bf q}\, \gamma_0
\,  {\cal P}_{\bf q}^s \, 
\Lambda_{\bf q}^{-e'}\, \gamma_\n \, T_a\, {\cal M}^\dagger_{\bf k} \, 
\Lambda_{\bf k}^e\right]}{{\rm Tr}
\left[{\cal M}_{\bf k}\,  {\cal M}^\dagger_{\bf k} \, 
\Lambda_{\bf k}^e\right]} \,\, .
\ee
The form (\ref{gap3}) of the gap equation holds for all cases considered 
in this paper. What is different in each case is the structure
of the term ${\cal T}_{\m\n}^{ee',s}({\bf k},{\bf q})$. 
Our computation will be done in pure Coulomb gauge, where
\be
\Delta^{00}(P)=\Delta_\ell(P) \,\, , \,\,\, \Delta^{0i}(P)=0 \,\, ,
\,\,\,  \Delta^{ij}(P)=(\delta^{ij}-\hat{p}^i\hat{p}^j)\, \Delta_t(P)
\,\, ,
\ee
with the longitudinal and transverse propagators $\Delta_{\ell,t}$ and
$P\equiv K-Q$.
Consequently, we only need the 00-component, 
${\cal T}_{00}^{ee',s}({\bf k},{\bf q})$, and the transverse 
projection of the $ij$-components, 
\be
{\cal T}_t^{ee',s}({\bf k},{\bf q})\equiv -(\delta^{ij}-\hat{p}^i\hat{p}^j) 
\, {\cal T}_{ij}^{ee',s}({\bf k},{\bf q}) \,\, ,
\ee
of the tensor (\ref{T2SC}). (The extra minus sign is included for
the sake of notational convenience.)
It will turn out that in all cases studied here
the quantities ${\cal T}_{00,t}^{ee',s}({\bf k},{\bf q})$
are related in the following way:
\be \label{relation2}
\frac{{\cal T}_{00}^{ee',2}({\bf k},{\bf q})}{
{\cal T}_{00}^{ee',1}({\bf k},{\bf q})} 
= \frac{{\cal T}_t^{ee',2}({\bf k},{\bf q})}{
{\cal T}_t^{ee',1}({\bf k},{\bf q})} 
= \mbox{const.} \,\, .
\ee

The right-hand side of Eq.\ (\ref{T2SC}) depends on $k$, $q$, and 
$\uk\cdot\uq$. The latter can be replaced by the square of the 
gluon 3-momentum $p^2$ via $\uk\cdot\uq=(k^2+q^2-p^2)/(2kq)$. 
Thus, the relevant components 
can be written in terms of a power series in $p^2$,
\begin{mathletters} \label{T}
\bea
{\cal T}_{00}^{ee',s}({\bf k},{\bf q})&=& a_s \sum_{m=-1}^\infty\,
\eta_{2m}^\ell(ee',k,q)\, \left(\frac{p^{2}}{kq}\right)^m \,\, ,\\
{\cal T}_t^{ee',s}({\bf k},{\bf q})
&=& a_s\sum_{m=-1}^\infty\,
\eta_{2m}^t(ee',k,q)\, \left(\frac{p^{2}}{kq}\right)^m\,\, .
\eea
\end{mathletters}
Here, the coefficients $\eta_{2m}^{\ell, t}(ee',k,q)$ no longer
depend on $s$ on account of Eq.\ (\ref{relation2}).
The overall normalization on the right-hand side of Eq.\ (\ref{T})
is still free, and we choose it such that Eq.\ (\ref{constraint})
is fulfilled. This 
uniquely determines the values of the dimensionless coefficients 
$\eta_{2m}^{\ell,t}(ee',k,q)$.

We now perform the Matsubara sum in Eq.\ (\ref{gap3}), which does
not depend on the detailed structure of the tensor 
${\cal T}_{\m\n}^{ee',s}({\bf k},{\bf q})$. This calculation is 
similar to that of Ref.\ \cite{rischke1}. The difference is the appearance
of the wave function renormalization factor $Z(q_0)$ \cite{wang}.
To subleading order, this amounts to an extra factor $Z(\e_{q,s}^{e'})$
in the gap equation. Since there are two different excitation energies 
$\e_{q,1}$ and $\e_{q,2}$ on the right-hand side of the gap equation, 
we can put the gap function on the left-hand side on either one of the 
two possible quasiparticle mass shells $k_0=\e_{k,1}$ or $k_0=\e_{k,2}$.
One then obtains
\begin{eqnarray}
\phi^e(\e_{k,r}^e,k) & = & \frac{g^2}{16\pi ^2 k} 
\int_{\mu-\delta}^{\mu+\delta}
dq \, q \sum_{e',s} a_s\,  Z(\e_{q,s}^{e'})\,
\frac{\phi^{e'}(\e_{q,s}^{e'},q)}{\e_{q,s}^{e'}}\,
\tanh\left(\frac{\e_{q,s}^{e'}}{2T}\right)\sum_m  \int_{|k-q|}^{k+q} dp\,p
 \left(\frac{p^2}{kq}\right)^m\, \left\{\frac{2}{p^2+3m_g^2}\,
\eta_{2m}^\ell   \right.
\non
&&+\left.\left[\;\frac{2}{p^2}\, \Theta(p-M)
+\Theta(M-p)\left(\frac{p^4}{p^6+M^4(\e_{q,s}^{e'}+\e_{k,r}^e)^2}
+\frac{p^4}{p^6+M^4(\e_{q,s}^{e'}-\e_{k,r}^e)^2}\right)\right]
\eta_{2m}^t \right\} \,\,.
\label{a1}
\end{eqnarray}
The first term in braces arises from static 
electric gluons, while the two terms in brackets originate 
from non-static and almost static magnetic gluons, respectively. Various 
other terms which yield sub-subleading contributions to the gap equation 
\cite{rischke1} have been omitted. 
In deriving Eq.\ (\ref{a1}) we assumed that the
gap function does not depend on the direction of ${\bf k}$. 
This is true in all cases considered here, except for the polar
phase, where we neglect this dependence, cf.\ Sec.\ \ref{polarphase}.

Although the coefficients $\eta_{2m}^{\ell,t}$ depend on $k$ and $q$,
to subleading order in the gap equation we may approximate 
$k\simeq q\simeq \m$. This can be easily proven by power counting.
To this end, it is sufficient to take $k= \m$, 
and write $q = \m + \xi$, where $\xi = q - \m$. 
In weak coupling, the gap function is sharply peaked around the
Fermi surface, and thus the range of integration
in the gap equation can be restricted to a small region of size $2\, \d$
around the Fermi surface. All that is necessary is that
$\d$ is parametrically much larger than $\phi_0$,
but still much smaller than $\m$, $\phi_0 \ll \d \ll \m$
\cite{rischke1}. It turns out that $\d \sim m_g$ is a convenient choice.
Since the integral over $\x$ is symmetric around $\x = 0$, 
terms proportional to odd powers of $\x$ vanish by symmetry.
Thus, corrections to the leading-order terms are at most $\sim
(\x/\m)^2$. As long as $\d$ is parametrically of the order of $m_g$,
$\x \leq m_g$, and these corrections are $\sim g^2$,
i.e., suppressed by two powers of the coupling
constant. Even for the leading terms in the gap equation the
correction due to terms 
$\sim (\x/\m)^2$ is then only of sub-subleading order and thus
negligible.

Since the coefficients $\eta_{2m}^{\ell,t}$ are dimensionless, 
with the approximation $k \simeq q \simeq \m$ they become pure 
numbers which, as we shall see in the following, are directly 
related to the constant $d$ discussed in
the introduction and listed in Table \ref{table1}. In all cases considered
here, $\eta_{2m}^{\ell,t}=0$ for 
$m\geq 3$, and the series in Eq.\ (\ref{T}) terminate after the first few
terms. Moreover, $\eta_{-2}^\ell$ always vanishes and, to subleading order, 
also $\eta_{-2}^{t}=0$.
For the remaining $m$, the $p$ integral in Eq.\ (\ref{a1}) 
can be performed exactly. 
The details of this calculation are deferred to Appendix \ref{AppB}.
We obtain
\begin{eqnarray}
\phi^e(\e_{k,r}^e,k) & = & \frac{g^2}{16\pi ^2} 
\int_{\mu-\delta}^{\mu+\delta}
dq \sum_{e',s} a_s \, Z(\e_{q,s}^{e'})\,
\frac{\phi^{e'}(\e_{q,s}^{e'},q)}{\e_{q,s}^{e'}}\,
\tanh\left(\frac{\e_{q,s}^{e'}}{2T}\right) \non 
&&\times \left[\eta_0^t\, \frac{1}{3}\, 
\ln\frac{M^2}{|(\e_{q,s}^{e'})^2-(\e_{k,r}^e)^2|}
+\eta_0^\ell\, \ln\frac{4\mu^2}{3m_g^2}
+\eta_0^t\, \ln\frac{4\mu^2}{M^2}
+4(\eta_2^\ell+\eta_2^t)+8(\eta_4^\ell+\eta_4^t)\right] \; .
\label{a1-1}
\end{eqnarray}
Note that the contribution from almost static magnetic gluons 
only appears in the term  proportional to $\eta_0^t$, 
while non-static magnetic and static electric gluons 
contribute to all other terms.

The antiparticle contribution ($e'=-$)
does not have a BCS logarithm, since $\e_{q,s}^-\simeq q+\m$. 
For the same reason, for antiparticles the logarithm from almost static
magnetic gluons is also only of order 1,
and furthermore there is no large logarithm from the $p$ integrals.
Therefore, the antiparticles contribute at most to sub-subleading
order to the gap equation and can be neglected. In the following,
we may thus set $e=e'=+$ and omit this superscript for the sake
of simplicity. Then, the gap equation for the quasiparticle gap 
function reads 
\begin{equation} \label{gap4}
\phi(\e_{k,r},k)=\bar{g}^2
\int_0^\delta d(q-\mu)\; \sum_s a_s \, Z(\e_{q,s})\,
\frac{\phi(\e_{q,s},q)}{\e_{q,s}}\;
\tanh\left(\frac{\e_{q,s}}{2T}\right)\;\frac{3}{4}\, \eta_0^t\,
\ln\left(\frac{b^2\mu^2}{|\e_{q,s}^2-\e_{k,r}^2|}\right) \,\, ,
\end{equation}
where
\be
b^2=\frac{64\, \m^4}{M^4}\left(\frac{4\m^2}{3m_g^2}
\right)^{3\eta_0^\ell/\eta_0^t}\exp(-2d) \,\, ,
\ee
with
\be \label{d}
 d=-\frac{6}{\eta_0^t}\, \left[\eta_2^\ell+\eta_2^t+2(\eta_4^\ell+\eta_4^t)
\right] \,\, .
\ee
In all cases considered in this paper, $\eta_0^\ell=\eta_0^t$, so that
$b$ assumes the value quoted in Eq.\ (\ref{b}). 
The expression (\ref{d}) is a general formula to compute the 
constant $d$ from the coefficients $\eta_{2m}^{\ell,t}$.
We also find that, for all cases considered here, 
$\eta_0^t=2/3$. This is the uniqueness of the leading-order
contribution
to the gap equation mentioned before. With this value
of $\eta_0^t$, the gap equation has the general
form (\ref{gap1}).

In the following, we shall discuss spin-zero color superconductors in 
the 2SC and CFL phases, as well as 
spin-one color superconductors in the CSL phase
with both longitudinal and transverse gaps,
the CSL phase with longitudinal gap only and with transverse gap 
only, and the polar phase. 
Each case is uniquely characterized by the matrix 
${\cal M}_{\bf k}$ which is given by the symmetries of the 
color-superconducting condensate.
This matrix determines the eigenvalues $\lambda_r$ and 
the projectors ${\cal P}_{\bf k}^r$. Evaluating the traces in Eq.\ 
(\ref{T2SC}) and comparing with Eq.\ (\ref{T}), one reads off the 
coefficients $\eta_{2m}^{\ell,t}$, as well as the constants $a_r$. 
This completely specifies the gap equation in each case.

\subsection{The 2SC phase} \label{2SC}

For $N_f=2$, the spin-zero condensate is a singlet in flavor and an 
antitriplet in color space \cite{bailin}. The (antisymmetric) singlet 
structure in flavor space can be represented by the second 
Pauli matrix $(\tau_2)_{fg}=i\e_{fg}$, $f,g=1,2$. The 
(antisymmetric) antitriplet
structure in color space restricts the gap matrix to be
a linear combination 
of the antisymmetric Gell-Mann matrices $\lambda_2$, $\lambda_5$,
and $\lambda_7$. These form an $SO(3)$ subgroup of $SU(3)_c$,
so that we can also choose the generators $(J_i)_{jk}=-i\e_{ijk}$, 
$i,j,k=1,2,3$, of $SO(3)$. The gap matrix is thus a scalar 
in flavor space 
and a  3-vector in color space. Upon condensation, this vector points in 
an arbitrary, but fixed, direction which breaks $SU(3)_c$ to $SU(2)_c$.
For the sake of convenience, 
we align this vector with $J_3$. Thus, the matrix ${\cal M}_{\bf k}$ reads
\be \label{M2SC}
{\cal M}_{\bf k}=J_3\,\tau_2 \, \gamma_5 \,\, ,
\ee
where $\gamma_5$ takes into account that we restrict our discussion
to the even-parity channel. This matrix obviously fulfills the condition 
(\ref{M}). From Eq.\ (\ref{Ldef}) we construct the matrix
\be \label{L2SC}
\left(L_{\bf k}\right)_{ij}^{fg} = (J_3^2)_{ij} \, (\tau_2^2)^{fg} = (\d_{ij}-
\d_{i3}\d_{j3})\, \d^{fg} \,\, .
\ee
In this case, $L_{\bf k}$ does not depend on ${\bf k}$, and consists of
a unit matrix in flavor space and a projector onto the first two colors
in color space. In principle, it also consists of a unit matrix in
Dirac space, which we disregard on account of the spin-zero nature of
the condensate.

The eigenvalues of $L_{\bf k}$ are (cf.\ Appendix \ref{AppA})
\be \label{EV2SC}
\lambda_1=1 \quad (\mbox{4-fold}) \quad , \qquad \lambda_2=0 \quad 
(\mbox{2-fold}) \,\, .
\ee
{}From Eq.\ (\ref{excite2}) we conclude that there are four gapped and
two ungapped excitations. 

The projectors ${\cal P}_{\bf k}$ follow from Eq.\ (\ref{proj}),
\be \label{p2SC}
{\cal P}_{\bf k}^1=L_{\bf k} \quad , \qquad {\cal P}_{\bf k}^2=1-L_{\bf k}
\,\, .
\ee
They have the property that $J_3{\cal P}_{\bf k}^1=J_3$ and
$J_3{\cal P}_{\bf k}^2=0$. Consequently, the tensor 
${\cal T}_{\m\n}^{ee',2}({\bf k},{\bf q})$ vanishes trivially.
For $s=1$ we obtain
\begin{mathletters} \label{T2SC2}
\bea
{\cal T}_{00}^{ee',1}({\bf k},{\bf q})&=&\frac{1}{3}\,\left(1+
ee'\,\uk\cdot\uq\right) \,\, ,\\
{\cal T}_t^{ee',1}({\bf k},{\bf q})
&=&\frac{1}{3}\, \left[3-ee'\,\uk\cdot\uq-\frac{(ek-e'q)^2}{p^2}\,
\left(1+ee'\,\uk\cdot\uq\right)\right] \,\, .
\eea
\end{mathletters}
We now match this result to the expansion in terms of $p^2$, Eq.\ (\ref{T}).
Since ${\cal T}_{\m\n}^{ee',2}({\bf k},{\bf q})=0$
and because of Eq.\ (\ref{constraint}), we have 
\be \label{a2SC}
a_1=1 \quad , \qquad a_2=0 \,\, .
\ee
This uniquely fixes the coefficients $\eta_{2m}^{\ell,t}(ee',k,q)$. 
To subleading order we only require their values for
$e=e'=+$ and $k\simeq q\simeq \m$, 
\be \label{eta2SC}
\eta_0^\ell=\frac{2}{3} \quad , \qquad 
\eta_2^\ell=-\frac{1}{6} \quad , \qquad 
\eta_4^\ell=0 \quad , \qquad
\eta_0^t=\frac{2}{3} \quad, \qquad  
\eta_2^t=\frac{1}{6} \quad , \qquad 
\eta_4^t=0 \,\, .
\ee
This result implies that the contributions from static electric and 
non-static magnetic gluons to the constant $d$ defined in Eq.\ (\ref{d})
cancel, and consequently $d=0$.

\subsection{The CFL phase} \label{CFL} 
 
In the CFL phase, the spin-zero condensate is 
a flavor antitriplet locked with 
a color antitriplet \cite{alford2}, 
\be \label{MCFL}
{\cal M}_{\bf k}={\bf J} \cdot {\bf I} \, \gamma_5 \,\, ,
\ee
where ${\bf J}=(J_1,J_2,J_3)$ represents the antitriplet in color space,
with $(J_i)_{jk}=-i\e_{ijk}$ as introduced above. 
The vector ${\bf I}$ represents
the antitriplet in flavor space and is defined analogously. Consequently,
$({\bf J}\cdot{\bf I})_{ij}^{fg}=-\delta_i^f\,\delta_j^g+\delta_i^g\,
\delta_j^f$. This condensate breaks $SU(3)_c \times SU(3)_f$ to 
$SU(3)_{c+f}$.

{}From Eq.\ (\ref{Ldef}) we obtain the matrix
\be
(L_{\bf k})_{ij}^{fg}=\left[({\bf J} \cdot {\bf I})^2\right]_{ij}^{fg}
=\delta_i^f\,\delta_j^g+\delta_{ij}\,\delta^{fg} \,\, .
\ee
As in the 2SC case, the operator $L_{\bf k}$ is independent of 
${\bf k}$, and we omitted its trivial Dirac structure. 
It can be expanded in terms of its eigenvalues and 
projectors as in Eq.\ (\ref{L}), with (cf.\ Appendix \ref{AppA})
\be \label{EVCFL}
\lambda_1=4 \quad (\mbox{1-fold}) \quad , \qquad \lambda_2=1 \quad 
(\mbox{8-fold}) \,\, ,
\ee
and 
\be
({\cal P}_{\bf k}^1)_{ij}^{fg}=\frac{1}{3}\,\delta_i^f\,\delta_j^g
\quad , \qquad 
({\cal P}_{\bf k}^2)_{ij}^{fg}=\delta_{ij}\delta^{fg}-\frac{1}{3}\,
\delta_i^f\,\delta_j^g
\,\, ,
\ee
where ${\cal P}_{\bf k}^1$ and ${\cal P}_{\bf k}^2$ correspond to
the singlet and octet projector introduced in Ref.\ \cite{zarembo}.

We now compute the relevant components of the tensor  
${\cal T}_{\m\n}^{ee',s}({\bf k},{\bf q})$. 
Since the Dirac structure of ${\cal M}_{\bf k}$ 
is the same as in the 2SC case, the dependence
on ${\bf k}$ and ${\bf q}$ is identical to the one in Eq.\ (\ref{T2SC2}). 
However, since the color-flavor structure is different, 
we obtain a non-trivial
result both for $s=1$ and $s=2$, with different prefactors,
\begin{mathletters} \label{TCFL}
\bea
{\cal T}_{00}^{ee',1}({\bf k},{\bf q})=
\frac{1}{2} \, {\cal T}_{00}^{ee',2}({\bf k},{\bf q})&=&\frac{1}{9}\,\left(1+
ee'\,\uk\cdot\uq\right) \,\, ,\\
{\cal T}_t^{ee',1}({\bf k},{\bf q})=  
\frac{1}{2} \, {\cal T}_t^{ee',2}({\bf k},{\bf q})
&=&\frac{1}{9}\, \left[3-ee'\,\uk\cdot\uq-\frac{(ek-e'q)^2}{p^2}\,
\left(1+ee'\,\uk\cdot\uq\right)\right] \,\, .
\eea
\end{mathletters}
Obviously, the condition (\ref{relation2}) is fulfilled.
The coefficients $\eta_{2m}^{\ell,t}$ remain the same as 
in Eq.\ (\ref{eta2SC}), which again yields $d=0$. However, the
two-gap structure leads to the constants 
\be \label{aCFL}
a_1=\frac{1}{3} \quad , \qquad a_2=\frac{2}{3} \,\, .
\ee

In our treatment we have so far neglected the color-sextet,
flavor-sextet gap which is induced by condensation in the 
color-antitriplet, flavor-antitriplet channel \cite{pisarski}. 
Such a color-flavor symmetric structure is generated 
in the anomalous propagator $S_{21}$, even for the
completely antisymmetric order parameter of Eq.\ (\ref{MCFL}).
(This does not happen in the 2SC case, where the color-flavor structure
of $S_{21}$ remains completely antisymmetric.)
Consequently, it also appears on the right-hand side of the gap equation.    
The reason why it disappeared in our calculation is that we projected 
exclusively onto the antisymmetric color-flavor channel when we multiplied 
both sides of Eq.\ (\ref{gap2}) with 
${\cal M}^\dagger_{\bf k}\,\Lambda_{\bf k}^e$ and traced over color, 
flavor, and Dirac space.
To be consistent, one should have started with an order parameter
which includes both the symmetric and the antisymmetric color-flavor
structures. In weak coupling, however, the symmetric gap is suppressed by
an extra power of the strong coupling constant $g$ \cite{schaefer2}.
This fact by itself is not sufficient to neglect the symmetric gap in
the weak-coupling solution of the gap equation because, as explained
in the introduction, this could still lead to a subleading correction
which modifies the prefactor of the (antisymmetric) gap.
One way to avoid this is a cancellation
of the leading terms involving the symmetric gap in the gap
equation for the antisymmetric gap. A more detailed investigation
of this problem, however, is outside the scope of the present paper.

\subsection{The CSL phase} \label{CSLphase}

For condensation in the even-parity, spin-one channel the gap matrix
reads (cf.\ Appendix \ref{AppD}, see also Ref.\ \cite{rischke1}) 
\be
\Sigma_{21}(K)=\sum_{e=\pm}\bm{\phi}^e (K)
\cdot \big[\uk+\gperp({\bf k})\big]\; 
\Lambda^e_{\bf k} \;, 
\label{gmspin1} 
\ee 
where $\gperp (\mathbf{k})\equiv\vg-\vg\cdot\uk\,\uk $ and 
$\vg =(\g^1,\g^2,\g^3)$. This most general form for the gap matrix
in the spin-one case differs from the one in Ref.\ \cite{schaefer} by
the appearance of $\gperp({\bf k})$ instead of $\vg$. From the discussion in
Appendix \ref{AppD} it is obvious that both forms are equivalent. 

The spin-one condensate is an $SU(2)$ triplet, and thus 
the order parameter $\bm{\phi}^e (K)$ is a 3-vector. 
In the CSL phase, each spatial component of this vector 
is assigned a direction in color space, $(x,y,z)\rightarrow (r,g,b)$.
This breaks color $SU(3)_c$ and spatial $SO(3)$ to an
$SO(3)$ subgroup of joint color and spatial rotations. The 
matrix ${\cal M}_{\bf k}$ reads
\be \label{MCSL}
{\cal M}_{\bf k}={\bf J}\cdot \big[\uk+\gperp({\bf k})\big] \,\, .
\ee
This matrix fulfills the condition (\ref{M}), due to the fact
that $\Lambda_{\bf k}^e$ commutes with 
$\gperp({\bf k})$. Had we used $\vg$ in Eq.\ (\ref{gmspin1}), like
in Ref.\ \cite{schaefer}, this condition would have been violated
and the general discussion presented in Sec.\ \ref{generalder} 
would not apply to the subsequent calculation. 

{}From Eqs.\ (\ref{Ldef}) and (\ref{MCSL}) we compute
\be \label{LCSL}
(L_{\bf k})_{\alpha\beta}^{ij}=\delta^{ij}\, \delta_{\alpha\beta} +
\left[\hat{k}^i\, \d_{\alpha\gamma}+\gamma_{\perp\a\g}^i({\bf k})\right]
\left[\hat{k}^j\, \d_{\gamma\b}-\gamma_{\perp\g\b}^j({\bf k})\right] \,\, .
\ee
In contrast to the 2SC and CFL cases, this $12\times 12$ matrix in 
color and Dirac space now explicitly depends on ${\bf k}$. Nevertheless,
its eigenvalues are pure numbers (cf.\ Appendix \ref{AppA}),
\be \label{EVCSL}
\lambda_1=4 \quad (\mbox{4-fold}) \quad , \qquad \lambda_2=1 \quad 
(\mbox{8-fold}) \,\, . 
\ee
The projectors follow from Eq.\ (\ref{proj}),
\begin{mathletters}
\bea
({\cal P}^1_{\bf k})_{\alpha\beta}^{ij}&=&\frac{1}{3} \, 
\left[\hat{k}^i\, \d_{\alpha\gamma}+\gamma_{\perp\a\g}^i({\bf k})\right]
\left[\hat{k}^j\, \d_{\gamma\b}-\gamma_{\perp\g\b}^j({\bf k})\right] 
\,\, ,\\
({\cal P}^2_{\bf k})_{\alpha\beta}^{ij}&=&\delta^{ij}\, \delta_{\alpha\beta} -
\frac{1}{3} \,
\left[\hat{k}^i\, \d_{\alpha\gamma}+\gamma_{\perp\a\g}^i({\bf k})\right]
\left[\hat{k}^j\, \d_{\gamma\b}-\gamma_{\perp\g\b}^j({\bf k})\right] \,\, .
\eea
\end{mathletters}
Inserting these projectors and ${\cal M}_{\bf k}$ from Eq.\ (\ref{MCSL})
into Eq.\ (\ref{T2SC}) we obtain
\begin{mathletters} \label{TCSL}
\bea
\frac{1}{2}\,{\cal T}_{00}^{ee',1}({\bf k},{\bf q})&=&
{\cal T}_{00}^{ee',2}({\bf k},{\bf q})=
\frac{1}{27} \, \left(1+ee'\,\uk\cdot\uq\right)
\left[1+(1+ee')\,\uk\cdot\uq\right] \,\, ,\\
\frac{1}{2}\,{\cal T}_t^{ee',1}({\bf k},{\bf q})&=&
{\cal T}_t^{ee',2}({\bf k},{\bf q})=
\frac{1}{27} \, \left\{  2 \, \uk \cdot \uq \left( 1 - e e'
\, \uk \cdot \uq \right) \right. \\
&& \hspace*{2.5cm}\left.  + \left[ 1- \frac{(ek-e'q)^2}{p^2} \right]\,
\left(1+ee'\,\uk\cdot\uq\right)\left[1+(1+ee')\,
\uk\cdot\uq\right]\right\} \,\, .
\eea
\end{mathletters}
Comparing this to Eq.\ (\ref{TCFL}), the prefactor 1/2 now accompanies
${\cal T}_{00,t}^{ee',1}$ instead of ${\cal T}_{00,t}^{ee',2}$.
Consequently, the constants $a_1$ and $a_2$ exchange their
roles compared to the CFL case, Eq.\ (\ref{aCFL}),
\be  \label{aCSL}
a_1=\frac{2}{3} \quad , \qquad a_2=\frac{1}{3}  
\ee
and, to subleading order,
\be \label{etaCSL}
\eta_0^\ell=\frac{2}{3} \quad , \qquad 
\eta_2^\ell=-\frac{7}{18} \quad , \qquad 
\eta_4^\ell=\frac{1}{18} \quad , \qquad 
\eta_0^t=\frac{2}{3} \quad , \qquad 
\eta_2^t=-\frac{5}{18} \quad , \qquad 
\eta_4^t=0 \,\, .
\ee
According to Eq.\ (\ref{d}), this yields $d=5$. 

As in the CFL case, another condensate 
with a symmetric color structure is
induced. This condensate belongs to the color-sextet representation and, 
for $N_f=1$, necessarily carries spin zero. To identify this induced 
condensate, one has to explicitly 
analyze the color structure of $S_{21}$. By analogy to the CFL case, we 
expect this condensate to be suppressed by a power of $g$ compared
to the primary spin-one, color-antitriplet condensate. 
Its contribution to the gap equation could be of sub-subleading order,
if there is a cancellation of the leading terms involving the
spin-zero gap in the gap equation for the spin-one gap. A more
detailed investigation, however, is beyond the scope of the 
present paper.

In the following two subsections we study two special cases of the CSL
color superconductor. The first is 
${\cal M}_{\bf k}\sim {\bf J}\cdot\hat{\bf k}$, and the second is 
${\cal M}_{\bf k}\sim {\bf J}\cdot\gperp({\bf k})$.
In the first case, the gap matrix commutes with the chirality
projector ${\cal P}_{r,\ell}=(1\pm\g_5)/2$, and
consequently only quarks of the same chirality form Cooper pairs.
The ensuing gap was termed {\it longitudinal gap\/} in Ref.\ \cite{rischke1}.
(It corresponds to the LL and RR gaps of Ref.\ \cite{schaefer}.)
In the second case, commuting the gap matrix with the 
chirality projector flips the sign of chirality,
which indicates that the quarks in the Cooper pair have opposite chirality.
This leads to the so-called {\it transverse gap\/} 
\cite{rischke1} (the LR and RL gaps of Ref.\ \cite{schaefer}).
The reason why we study both cases separately is that a purely 
longitudinal gap matrix on the right-hand side does not induce 
a transverse gap on the left-hand side of the gap equation and
vice versa. This will be explained in more detail below.

\subsection{The longitudinal CSL phase} \label{longphase}

In the CSL phase with longitudinal gaps only, the matrix 
${\cal M}_{\bf k}$ reads
\be \label{MCSLlong}
{\cal M}_{\bf k}={\bf J}\cdot\uk \,\, .
\ee
The condition (\ref{M}) is trivially fulfilled. Inserting Eq.\ 
(\ref{MCSLlong}) into Eq.\ (\ref{Ldef}), we obtain
\be \label{LCSLlong}
(L_{\bf k})_{\a\b}^{ij}=\left(\d^{ij}-\hat{k}^i\,\hat{k}^j\right)\, \d_{\a\b} 
\,\, .
\ee
This matrix is a projector onto the subspace orthogonal to $\hat{\bf k}$.
However, due to color-spin locking, the indices $i,j$ run over  
fundamental colors and not over spatial dimensions, and thus, amusingly, 
this projection actually occurs in color space. 
Since $L_{\bf k}$ is a projector, we find the eigenvalues
(cf.\ Appendix \ref{AppA})
\be \label{EVlong}
\lambda_1=1 \quad (\mbox{8-fold}) \quad , \qquad \lambda_2=0 \quad 
(\mbox{4-fold}) \,\, . 
\ee
The projectors ${\cal P}_{\bf k}^{1,2}$ follow from Eq.\ (\ref{proj}),
\be \label{pCSLlong}
{\cal P}_{\bf k}^1=L_{\bf k} \quad , \qquad {\cal P}_{\bf k}^2=1-L_{\bf k}
\,\, ,
\ee
similar to the 2SC case, cf.\ Eq.\ (\ref{p2SC}).
The peculiar feature of Eq.\ (\ref{pCSLlong}) is that the projector 
${\cal P}_{\bf k}^1$ belongs to the eigenvalue corresponding to 
quasiparticle excitations with a longitudinal gap, but it actually 
projects onto the subspace orthogonal to $\hat{\bf k}$. This is,
however, not a contradiction, since the projection occurs in color
space, while the gap is longitudinal (parallel to $\hat{\bf k}$)
in real space.

The similarity to the 2SC case carries over to the quantities
${\cal T}_{00,t}^{ee',s}({\bf k},{\bf q})$. For $s=2$, 
these quantities again vanish
because ${\bf J}\cdot\uk \, {\cal P}_{\bf k}^2 =0$.
For $s=1$, we obtain
\begin{mathletters} \label{TCSLlong}
\bea
{\cal T}_{00}^{ee',1}({\bf k},{\bf q})&=&\frac{1}{3}\,\uk\cdot\uq \, \left(1+
ee'\,\uk\cdot\uq\right) \,\, ,\\
{\cal T}_t^{ee',1}({\bf k},{\bf q})
&=&\frac{1}{3}\,\uk\cdot\uq\, \left[3-ee'\,\uk\cdot\uq-\frac{(ek-e'q)^2}{p^2}\,
\left(1+ee'\,\uk\cdot\uq\right)\right] \,\, ,
\eea
\end{mathletters}
which only differ by an overall factor $\uk\cdot\uq$ from those of
Eq.\ (\ref{T2SC2}). While the constants $a_r$ are the same as in the 2SC 
case, see Eq.\ (\ref{a2SC}), this factor substantially 
changes the coefficients $\eta_{2m}^{\ell,t}$,
\be \label{etaCSLlong}
\eta_0^\ell=\frac{2}{3} \quad , \qquad 
\eta_2^\ell=-\frac{1}{2} \quad , \qquad 
\eta_4^\ell=\frac{1}{12} \quad , \qquad 
\eta_0^t=\frac{2}{3} \quad , \qquad 
\eta_2^t=-\frac{1}{6} \quad , \qquad 
\eta_4^t=-\frac{1}{12} \,\, .
\ee
This leads to $d=6$.

We finally comment on why it is impossible that a purely longitudinal
order parameter induces a transverse gap. Inserting the matrix
${\cal M}_{\bf k}$ from Eq.\ (\ref{MCSLlong}) into the anomalous
propagator $S_{21}$ from Eq.\ (\ref{S212SC}), and the result into the 
right-hand side of the gap equation (\ref{gap2}), we realize that 
the resulting Dirac structure still commutes with $\g_5$ and thus preserves
the chirality. This is the characteristic feature of a longitudinal
gap. Therefore, the ansatz (\ref{MCSLlong}) does not induce 
a transverse gap on the right-hand side of the gap equation.

\subsection{The transverse CSL phase}  \label{transphase}

For transverse gaps,
\be \label{MCSLtrans}
{\cal M}_{\bf k}={\bf J}\cdot\gperp({\bf k}) \,\, .
\ee
The condition (\ref{M}) is fulfilled because $\gperp({\bf k})$
commutes with the energy projector $\Lambda_{\bf k}^e$. 
For the matrix $L_{\bf k}$ we obtain
\be \label{LCSLtrans}
(L_{\bf k})_{\a\b}^{ij}=2\,\hat{k}^i\,\hat{k}^j\, \d_{\a\b}
-\g_{\perp\a\g}^i({\bf k}) \, \g_{\perp\g\b}^j ({\bf k})
\,\, .
\ee
The eigenvalues of this matrix are (cf.\ Appendix \ref{AppA})
\be \label{EVtrans}
\lambda_1=2 \quad (\mbox{8-fold}) \quad , \qquad \lambda_2=0 \quad 
(\mbox{4-fold}) \,\, . 
\ee
The projectors ${\cal P}_{\bf k}^{1,2}$ are given by
\be \label{pCSLtrans}
{\cal P}_{\bf k}^1=\frac{1}{2}\,L_{\bf k} \quad , \qquad 
{\cal P}_{\bf k}^2=1-\frac{1}{2}\,L_{\bf k}
\,\, .
\ee
Although ${\bf J}\cdot\gperp({\bf k})\, {\cal P}_{\bf k}^s \neq 0$ for both 
$s=1$ and $s=2$, the final result for
${\cal T}_{00,t}^{ee',2}({\bf k},{\bf q})$ is nevertheless zero.
To see this, however, one has to explicitly perform the trace 
in Eq.\ (\ref{T2SC}). For $s=1$, we obtain
\begin{mathletters} \label{TCSLtrans}
\bea
{\cal T}_{00}^{ee',1}({\bf k},{\bf q})&=&\frac{1}{6} \, \left(1+
ee'\,\uk\cdot\uq\right)^2 \,\, ,\\
{\cal T}_t^{ee',1}({\bf k},{\bf q})
&=&\frac{1}{6}\, \left(1+ee'\,\uk\cdot\uq\right)^2 \, 
\left[1-\frac{(ek-e'q)^2}{p^2}\right] \,\, .
\eea
\end{mathletters}
The constants $a_r$ are the same as in the 2SC and longitudinal CSL phases, 
see Eq.\ (\ref{a2SC}). 
The coefficients $\eta_{2m}^{\ell,t}$ are
\be \label{etaCSLtrans}
\eta_0^\ell=\frac{2}{3} \quad , \qquad 
\eta_2^\ell=-\frac{1}{3} \quad , \qquad 
\eta_4^\ell=\frac{1}{24} \quad , \qquad 
\eta_0^t=\frac{2}{3} \quad , \qquad 
\eta_2^t=-\frac{1}{3} \quad , \qquad 
\eta_4^t=\frac{1}{24} \,\, .
\ee
This gives $d=9/2$.

For the same reasons as explained at the end of the last subsection,
it is impossible to induce a longitudinal gap with the 
matrix ${\cal M}_{\bf k}$ of Eq.\ (\ref{MCSLtrans}) on the
right-hand side of the gap equation.

\subsection{The polar phase} \label{polarphase}

In contrast to the CSL phase, in the polar phase  
the vector $\bm{\phi}^e(K)$ in Eq.\ (\ref{gmspin1}) does not couple
to color space. Instead, it
simply points into a fixed spatial direction, which we choose 
to be the $z$-axis.
Consequently, the matrix ${\cal M}_{\bf k}$ in Eq.\ (\ref{gm2SC}) reads
\be \label{Mpolar}
{\cal M}_{\bf k}=J_3 \, \left[\hat{k}^z+\gperp^z({\bf k})\right] \,\, .
\ee
As in the 2SC case, the condensate is aligned with the
(anti) blue direction in color space. Thus, condensation
spontaneously breaks the color $SU(3)_c$
and spatial $SO(3)$ symmetries to $SU(2)_c$ and $SO(2)$, respectively. 

Due to the identity
$(\hat{k}^z + \gamma_\perp^z)(\hat{k}^z - \gamma_\perp^z)=1$,
the Dirac structure of the matrix $L_{\bf k}$ is trivial, and it
looks rather similar as in the 2SC case, Eq.\ (\ref{L2SC}),
\be \label{Lpolar}
\left(L_{\bf k}\right)^{ij}_{\a\b} = (J_3^2)^{ij}\,\d_{\a\b}
 = (\d^{ij}-\d^{i3} \, \d^{j3}) \, \d_{\a\b} \,\, .
\ee
This similarity is also apparent in the eigenvalues of $L_{\bf k}$
(cf.\ Appendix \ref{AppA}),
\be  \label{EVpolar}
\lambda_1=1 \quad (\mbox{8-fold}) \quad , \qquad \lambda_2=0 \quad 
(\mbox{4-fold}) \,\, , 
\ee
where the degeneracy refers to the combined color and Dirac spaces.
The projectors are the same as in Eq.\ (\ref{p2SC}). For this reason,
we again immediately conclude that 
${\cal T}_{00,t}^{ee',2}({\bf k},{\bf q})=0$. For $s=1$ we obtain
\begin{mathletters} \label{Tpolar}
\bea
{\cal T}_{00}^{ee',1}({\bf k},{\bf q})&=&
\frac{1}{3}\,\left\{ 
\left( 1 + ee'\,\uk\cdot\uq \right)  \, 
\left[ 1 + (1+ee') \, \hat{k}^z \, \hat{q}^z \right]
 - (e\hat{k}^z + e'\hat{q}^z)^2 \right\} \,\, ,\\
{\cal T}_t^{ee',1}({\bf k},{\bf q})&=&
\frac{1}{3}\, \left(
2\, \hat{k}^z\, \hat{q}^z \, \left(1-e e' \uk \cdot \uq\right)
+ \left[ 1-\frac{(ek-e'q)^2}{p^2}\right] \right. \nonumber \\
&   & \left. \times 
\left\{ \left( 1+ee'\,\uk\cdot\uq \right) \left[ 1 + (1+ee')\,
\hat{k}^z \, \hat{q}^z\right]-(e\hat{k}^z+e'\hat{q}^z)^2
\right\} \right) \,\, .
\eea
\end{mathletters}
{}From this it is obvious that $a_1=1$ and $a_2=0$ as in the 2SC case,
cf.\ Eq.\ (\ref{a2SC}).

There is, however, a marked difference between the expressions 
(\ref{Tpolar}) and the 
corresponding ones for all previously discussed cases. 
In contrast to the other cases, there are two independent, fixed
spatial directions, that of the order parameter and that of the
vector ${\bf k}$. Since we 
already aligned the order parameter with the $z$-direction, we are
no longer free to choose ${\bf k}=(0,0,k)$ for the 
$d^3 {\bf q}$-integration. Without loss of generality, however,
we may assume ${\bf k}$ to lie in the $xz$-plane, i.e.,
${\bf k}=k\,(\sin\vartheta,0,\cos\vartheta)$, where $\vartheta$
is the angle between the order parameter and ${\bf k}$.  
In spherical coordinates for the $d^3{\bf q}$-integration,
the azimuthal angle $\theta$ 
is no longer identical with the angle between 
${\bf k}$ and ${\bf q}$. Or in other words, $\uk\cdot\uq$ no longer
depends solely on $\theta$, but also on the polar angle $\varphi$.
This has the consequence that also the modulus of the gluon 
3-momentum $p$ depends on $\varphi$. Since $p$ enters the gluon
spectral densities in a complicated fashion, it appears impossible
to perform the $\varphi$-integration analytically in this way. 

The solution is to rotate the coordinate frame for the 
$d^3{\bf q}$-integration by the angle $\vartheta$ around the $y$-axis,
such that the rotated $z$-direction aligns with ${\bf k}$.   
The quantities $\uk\cdot\uq$, $\hat{q}^z$, and $\hat{k}^z$
appearing in Eqs.\ (\ref{Tpolar}) are expressed in terms of the new 
spherical coordinates $(q,\theta',\varphi')$ and the rotation angle 
$\vartheta$ as follows:
\be 
\uk\cdot\uq=\cos\theta' \quad, \qquad \hat{q}^z=\cos\theta' \, \cos\vartheta 
-\sin\theta' \, \sin\vartheta \, \cos\varphi' \quad, \qquad 
\hat{k}^z=\cos\vartheta \,\, .
\ee 
In the new coordinates the angle between ${\bf k}$ and ${\bf q}$
is identical with the azimuthal angle $\theta'$, and thus
$p$ becomes independent of $\varphi'$. Still, the 
$\varphi'$-integral is not trivial because of the potential 
$\varphi'$ dependence of the gap function. At this point we can
only proceed by assuming the gap function to be independent of $\varphi'$.
With this assumption, the $\varphi'$-integration becomes elementary,
and we are finally able to read off the coefficients $\eta_{2m}^{\ell,t}$,
which now depend on $\vartheta$,
\be \label{etaCSLpolar}
\eta_0^\ell=\frac{2}{3} \; , \quad 
\eta_2^\ell=-\frac{2+\cos^2\vartheta}{6} \; , \quad 
\eta_4^\ell=\frac{1+\cos^2\vartheta}{24} \; , \quad 
\eta_0^t=\frac{2}{3} \; , \quad 
\eta_2^t=-\frac{2-\cos^2\vartheta}{6} \; , \quad 
\eta_4^t=\frac{1-3\cos^2\vartheta}{24} \,\, .
\ee
{}From this and Eq.\ (\ref{d}) we compute $d=3(3+\cos^2\vartheta)/2$.

Let us now comment on our assumption that the gap function is independent 
of $\varphi'$. As mentioned in the introduction
and as will be shown in the next section, the value
of the gap function at the Fermi surface, $\phi_0$, 
is proportional to $\exp(-d)$,
cf.\ also Table \ref{table1}. The angular dependence of $d$ then implies
a similar dependence of the gap itself. If ${\bf k}$ points in the same
direction as the order parameter, $\vartheta=0$, we find $d=6$, while for
${\bf k}$ being orthogonal to the order parameter, $\vartheta=\pi/2$,
one obtains $d=9/2$. In the first case, the gap is longitudinal in the sense
introduced in Sec.\ \ref{CSLphase}, while in the second it is 
transverse. These two cases have also been discussed in Refs.\ 
\cite{ren,schaefer}, with the same results for the constant $d$.
Our results surpass the previous ones in that they interpolate between these
two limiting cases. 

However, the angular dependence of $\phi_0$ causes the following problem.
The gap function $\phi(\e_{k,1},{\bf k})$ is proportional 
to $\phi_0$, cf.\ the 
next section, and thus also depends on $\vartheta$. Under
the $d^3{\bf q}$-integral on the right-hand side of the gap equation,
this dependence translates into a $\varphi'$ dependence of 
$\phi(\e_{q,1},{\bf q})$. Our previous assumption, which was necessary 
in order to perform the $\varphi'$-integral, precisely neglected this
dependence. Therefore, this approximation is in principle inconsistent.
Nevertheless, the agreement of our results with the ones of Refs.\
\cite{ren,schaefer} suggest that the 
$\varphi'$-dependence of the gap function could be a sub-subleading
effect.

\section{Solution of the gap equation} \label{solution}

In this section we solve the gap equation (\ref{gap1}).
Let us first distinguish between the cases where $a_1=1, \, a_2=0$, and
where both $a_1$ and $a_2$ are nonzero. 
The former are the 2SC phase, Sec.\ \ref{2SC},
the longitudinal and transverse CSL phases, Secs.\ \ref{longphase}
and \ref{transphase}, and the polar phase, Sec.\ \ref{polarphase}.
The latter are the CFL phase, Sec.\ \ref{CFL}, and the
CSL phase with both longitudinal and transverse gaps, Sec.\
\ref{CSLphase}.

In the former cases, there is only one gapped quasiparticle
excitation and the solution of the gap equation (\ref{gap1}) is
well-known. It was discussed in detail in Ref.\ \cite{wang}.
In the 2SC phase, the longitudinal CSL phase, and the polar phase,
all one has to do is replace the constant $\tilde{b}$
in the calculation of Ref.\ \cite{wang} by the constant
$b = \tilde{b} \exp(-d)$, cf.\ Eq.\ (\ref{b}). The result for
the value of the gap function at the Fermi surface is
Eq.\ (\ref{phi0}), but without the
factor $(\lambda_1^{a_1}\, \lambda_2^{a_2})^{-1/2}$.
However, one immediately reads off Table \ref{table1}
that in the respective cases this factor trivially equals one.
In the 2SC phase, $d=0$, and consequently $b= \tilde{b}$, such that
the result coincides with Eq.\ (\ref{phi02SC}). 
In the other phases, where $d >0$, the gap is reduced as compared
to the 2SC phase by a factor $\exp(-d)$, cf.\ Table \ref{table1}.

There is a slight subtlety when solving the gap equation
in the transverse CSL phase.
The value of the nonvanishing eigenvalue is $\lambda_1=2$, not 1.
One has to multiply both sides of Eq.\ (\ref{gap1}) with
$\sqrt{\lambda_1}$ in order to obtain a gap equation for which 
the solution of Ref.\ \cite{wang} applies.
This rescaling is 
appropriate, as in this case the gap in the quasiparticle excitation 
spectrum is indeed $2 \, \sqrt{\lambda_1}\, \phi_0$, 
and not simply $2\, \phi_0$. The factor
$(\lambda_1^{a_1}\, \lambda_2^{a_2})^{-1/2}$ in Eq.\ (\ref{phi0})
precisely accounts for this rescaling of the gap function, 
such that this equation is also valid in the transverse CSL phase.

In the CFL phase and the CSL phase with both longitudinal and
transverse gaps, there are two gapped quasiparticle excitations,
which renders the solution of Eq.\ (\ref{gap1}) somewhat more
complicated. 
A priori, one has to solve two gap equations, one 
for each quasiparticle mass shell, $k_0=\e_{k,1}$ and $k_0=\e_{k,2}$. 
Therefore, as a function of momentum $k$, there are in principle 
two different gap functions,
$\phi_r(k) \equiv \phi(\e_{k,r},k), \, r = 1,2$. 

In order to proceed with the solution, to subleading order 
we may approximate the logarithm in Eq.\ (\ref{gap1}) in a way 
first proposed by Son \cite{son},
\be
\frac{1}{2}\,\ln\left(\frac{b^2\m^2}{|\e_{q,s}^2-\e_{k,r}^2|}\right)
\simeq\Theta(\e_{q,s}-\e_{k,r})\, \ln \left(\frac{b\m}{\e_{q,s}}\right)
+\Theta(\e_{k,r}-\e_{q,s})\, \ln \left(\frac{b\m}{\e_{k,r}}\right) \,\, .
\ee
With this approximation and the new variables
\be \label{vartrans}
x_r \equiv \bar{g} \, \ln\left(\frac{2b\m}{k-\m+\e_{k,r}}\right) \quad, \qquad
y_s \equiv \bar{g} \, \ln\left(\frac{2b\m}{q-\m+\e_{q,s}}\right) \,\, ,
\ee
to subleading order the gap equation (\ref{gap1}) transforms into \cite{wang}
\be \label{phixr}
\phi(x_r)=\sum_s a_s \left\{ x_r \int_{x_r}^{x_s^*} dy_s  \, 
(1-2\,\bar{g}\,y_s) \, \tanh \left[\frac{\e(y_s)}{2T}\right]\, \phi(y_s)
 + \int_{x_0}^{x_r}dy_s \, y_s \, (1-2\, \bar{g}\,y_s) \, 
\tanh \left[\frac{\e(y_s)}{2T}\right]\, \phi(y_s)\right\} \,\, .
\ee
Here, we denoted the value of $x_s$ at the Fermi surface, i.e,
for $k= \mu$ and $\e_{k,s} = \e_{\mu,s}$, by
\be \label{xsdef}
x_s^* \equiv \bar{g} \, 
\ln\left(\frac{2b\m}{\sqrt{\lambda_s} \, \phi_{0,s}}\right) \,\, ,
\ee
where $\phi_{0,s} \equiv \phi(x_s^*)$ is the value of
the function $\phi(x_s)$ at the Fermi surface.
The single point $k= \mu$ in momentum space thus corresponds to two
different points $x_1^*,\, x_2^*$, $x_1^* \neq x_2^*$ ,
in the new variables $x_s$. 
Since we expect $\phi_{0,s}$ to be $\sim \exp(-1/\bar{g})$, 
$x_s^*$ is a constant of order one.  
Furthermore we defined
\be
x_0 \equiv \bar{g} \, \ln\left(\frac{b\m}{\d}\right) \,\, .
\ee
This constant is parametrically of order $O(\bar{g})$.
To subleading order, the relation between the new variable
$y_s$ and the excitation energy is given by \cite{rischke1},
\be
\e(y_s) = b\, \m \, \exp\left(-\frac{y_s}{\bar{g}}\right)\,\, .
\ee
A consequence of the transformation of variables (\ref{vartrans})
and of neglecting sub-subleading corrections is
that the two equations (\ref{phixr}) for $r=1$ and $r=2$ become
identical. The only difference is the notation for the argument of 
the function $\phi$, which in both cases we may simply call $x$.
Therefore, instead of two separate equations, we only have to 
consider a single equation which determines the function $\phi(x)$.
Moreover, $y_s$ is merely an integration variable, 
and we may set $y_s \equiv y$ in the following.

With Eq.\ (\ref{constraint}), we rewrite Eq.\ (\ref{phixr}) in the form
\begin{eqnarray}
\phi(x) & = & x \int_x^{x_2^*}dy\, (1- 2\, \bar{g} \, y) \,
\tanh\left[\frac{\e(y)}{2T}\right]  \, \phi(y)
+ \int_{x_0}^x dy\, y\, (1- 2\, \bar{g}\, y)\, 
\tanh \left[\frac{\e(y)}{2T}\right]\, \phi(y) \nonumber \\
&  & - a_1\, x \int_{x_1^*}^{x_2^*}dy\, (1- 2\, \bar{g} \, y) \,
\tanh\left[\frac{\e(y)}{2T}\right]\, \phi(y) \,\, .
\label{phix}
\end{eqnarray}
One can also write this equation in a form where
$x_2^*$ is replaced by $x_1^*$ and $a_1$ by $a_2$, respectively.
Equation (\ref{phix}) is an integral equation for the function 
$\phi(x)$, which is solved in the standard manner by converting 
it into a set of differential equations \cite{son},
\begin{mathletters}
\begin{eqnarray} \label{firstder}
\frac{d\phi}{dx} & = & \int_x^{x_2^*}dy\, (1- 2\, \bar{g}\, y)\,
\tanh \left[ \frac{\e(y)}{2T} \right]\, \phi(y)
- a_1 \int_{x_1^*}^{x_2^*} dy\,(1- 2\, \bar{g}\, y)\,
\tanh \left[ \frac{\e(y)}{2T} \right] \, \phi(y)\,\, ,  \\
\frac{d^2\phi}{dx^2} & = & - (1- 2\, \bar{g}\, x)\, 
\tanh \left[ \frac{\e(x)}{2T} \right]\, \phi(x) \, \, .
\label{diffeq}
\end{eqnarray}
\end{mathletters}
We now solve the second-order differential equation (\ref{diffeq}) at
zero temperature, $T=0$. One immediately observes that this equation is
identical to Eq.\ (22c) of Ref.\ \cite{wang}, and its solution
proceeds along the same lines as outlined there. The only difference
compared to the previous calculation are the extra terms 
$\sim a_1$ in Eqs.\ (\ref{phix}) and (\ref{firstder}).
To subleading order, we expect $\phi_{0,1}/ \phi_{0,2} \simeq 1$
(we show below that this assumption is consistent with our final result),
such that the difference
\be \label{x2x1}
x_2^* - x_1^* = \bar{g} \, \ln \left( 
\frac{\sqrt{\lambda_1}\, \phi_{0,1}}{\sqrt{\lambda_2}\, \phi_{0,2}} \right)
\simeq \frac{\bar{g}}{2}\, \ln \left( 
\frac{\lambda_1}{\lambda_2} \right)
\ee
is of order $O(\bar{g})$. Consequently, the extra terms $\sim a_1$
are of subleading order, $O(\bar{g}\phi_0)$, and we may
approximate
\be \label{subcorr}
\int_{x_1^*}^{x_2^*} dy\,(1- 2\, \bar{g}\, y)\,  \, \phi(y)
\simeq (x_2^*-x_1^*) \, \phi_{0,2} \,\, .
\ee
Since we always ordered the eigenvalues such that $\lambda_1 >
\lambda_2$, cf.\ Table \ref{table1}, $x_2^* - x_1^* >0$.

The subleading correction (\ref{subcorr}) qualitatively changes
the behavior of the gap function $\phi(x)$ near the Fermi surface.
In the absence of the term $\sim a_1$ in Eq.\ (\ref{firstder}), 
the derivative of the gap function vanishes for $x=x_2^*$,
and the gap function assumes its maximum at this point 
\cite{wang}. 
The subleading correction (\ref{subcorr}) induced by the two-gap
structure in the CFL and CSL phases
causes the derivative (\ref{firstder}) of the function
$\phi(x)$ to be {\it negative\/} at the Fermi surface. Consequently, since 
we still expect $\phi(x)$ to rapidly vanish away from the Fermi surface,
this function assumes its maximum not right {\it at\/} the Fermi surface,
but at a point $x_{\rm max}$ which is close, but not identical to
$x_2^*$. We shall see that $x_2^* - x_{\rm max} \sim O(\bar{g})$.

The subleading correction (\ref{subcorr}) modifies the solution 
of the differential equation (\ref{diffeq}) from the one
given in Ref.\ \cite{wang}. Again, we fix the two unknown constants in the
general solution of the second-order differential equation
(\ref{diffeq}) by matching the solution and its
derivative to the right-hand sides of Eqs.\ (\ref{phix}) and
(\ref{firstder}) at the point $x= x_2^*$.
Introducing the variables 
$z \equiv - (2 \bar{g})^{-2/3} \, (1-2\bar{g}x)$ and 
$z^* \equiv - (2 \bar{g})^{-2/3} \,(1-2\bar{g}x_2^*)$, 
the solution reads
\be \label{phiz}
\phi(z) = \phi_{0,2}  \left\{ \frac{M(|z|)}{M(|{z^*}|)} 
\frac{ \sin \left[\varphi(|z^*|) - \theta(|z|) \right] }{
\sin \left[\varphi(|z^*|) - \theta(|z^*|) \right] }
+ a_1 \, (x_2^*-x_1^*)\, (2\bar{g})^{-1/3} \,
\frac{M(|z|)}{N(|{z^*}|)}\,
\frac{ \sin \left[\theta(|z^*|) - \theta(|z|) \right] }{ 
\sin \left[\varphi(|z^*|) - \theta(|z^*|) \right] } \right\}\,\, ,
\ee
where the functions $M(|z|),\, N(|z|),\, \varphi(|z|),$ and 
$\theta(|z|)$ are related
to the Airy functions ${\rm Ai}(z), \, {\rm Bi}(z)$ and their derivatives
in the standard way \cite{abramowitz}.
The derivative $d\phi(z)/dz$ can be obtained from Eq.\ (\ref{phiz})
simply by replacing $M(|z|)$ and $\theta(|z|)$ by $N(|z|)$ and
$\varphi(|z|)$, respectively.
The difference to the solution for a single gapped quasiparticle
excitation, cf.\ Eq.\ (27) of Ref.\ \cite{wang}, 
is the term proportional to $a_1$.

Finally, we have to determine the value of $\phi_{0,2}$.
To this end, we rewrite Eq.\ (\ref{phix}) at the point $x= x_2^*$
in the form
\be \label{relation3}
\left[ z_0 + (2\bar{g})^{-2/3} \right] \, \frac{d \phi}{d z}(z_0) = 
\phi(z_0) \,\, ,
\ee
where $z_0 \equiv - (2 \bar{g})^{-2/3} \, (1-2\bar{g}x_0)$.
Remarkably, this equation holds in this form also in the case
of a single gapped quasiparticle excitation, 
cf.\ Eq.\ (29) of Ref.\  \cite{wang}.
In weak coupling, the dependence on the variable $z_0$ is spurious.
Inserting the solution (\ref{phiz}) and its derivative for 
$z=z_0$ and expanding 
$M(|z_0|),\, N(|z_0|),\, \varphi(|z_0|),$ and $\theta(|z_0|)$ 
to order $O(\bar{g})$ as demonstrated in Ref.\ \cite{wang},
one derives the condition
\be
x_2^* \simeq \frac{\pi}{2} + \bar{g} \, \frac{\pi^2+4}{8} + 
a_1\, (x_2^*-x_1^*)
\,\, .
\ee
The second term is the $O(\bar{g})$ correction originating from 
the quark self-energy. It leads to the constant $b_0'$ in
Eq.\ (\ref{phi02SC}) and was first derived in Refs.\ \cite{ren,wang}.
The last term $\sim a_1$ is the correction arising from the
two-gap structure in the CFL and CSL phases to the result (33) of
Ref.\ \cite{wang}.
Because of Eq.\ (\ref{x2x1}), this correction is also of order
$O(\bar{g})$.
Using the definition (\ref{xsdef}) of $x_2^*$, as well as the
condition (\ref{constraint}), we 
conclude that the expression for $\phi_{0,2}$ is identical to the
one for $\phi_0$ in Eq.\ (\ref{phi0}).
This is the value of the gap function at the Fermi surface, $k = \mu$,
or $x=x_2^*$, for the quasiparticle excitation branch $\e_{k,2}$. 
The additional suppression factor compared to
the 2SC gap $\phi_0^{\rm 2SC}$ of Eq.\ (\ref{phi02SC}),
which originates from the two-gap structure,
is $\left( \lambda_1^{a_1}\, \lambda_2^{a_2} \right)^{-1/2}$.
For the CFL phase, we obtain the value $2^{-1/3}$, while for
the CSL phase, we have $2^{-2/3}$.

We can also compute the gap function at the Fermi surface
for the first excitation branch $\e_{k,1}$, i.e., at $x=x_1^*$.
The difference $\phi_{0,2}- \phi_{0,1}$ can be obtained from
Eq.\ (\ref{phix}) as
\be
\phi_{0,2}- \phi_{0,1} = \int_{x_1^*}^{x_2^*} dy \, 
\left[ y-x_1^* - a_1\, (x_2^* - x_1^*) \right]
\, (1- 2\, \bar{g}\, y)\, \phi(y) \,\, .
\ee
An upper bound for the term in brackets is given by setting
$y = x_2^*$, where it assumes the value $a_2 (x_2^*-x_1^*)$
on account of Eq.\ (\ref{constraint}).
Pulling this factor out of the integral, the latter can
be estimated with Eq.\ (\ref{subcorr}). This proves that
the difference $\phi_{0,2}- \phi_{0,1}$ is only of order 
$O(\bar{g}^2\phi_0)$,
which shows that our above assumption
$\phi_{0,1}/\phi_{0,2} \simeq 1$ is consistent up to subleading order.
To this order, we may therefore set
$\phi_{0,1} = \phi_{0,2} \equiv \phi_0$.

We now determine the value of $x_{\rm max}$, 
where the gap function assumes its maximum, by
setting the left-hand side of Eq.\ (\ref{firstder}) equal to zero.
This leads to the condition
\be
\int_{x_{\rm max}}^{x_2^*}dy\, (1- 2\, \bar{g}\, y)\, \phi(y)
= a_1 \int_{x_1^*}^{x_2^*} dy\,(1- 2\, \bar{g}\, y)\, \phi(y)\,\, .
\ee
To order $O(\bar{g}\phi_0)$, one may easily solve this
equation for $x_{\rm max}$, with the result
\be
x_{\rm max} = x_2^* - a_1 \, \frac{\bar{g}}{2}\, 
\ln \left( \frac{\lambda_1}{\lambda_2} \right) \,\, ,
\ee
i.e., $x_{\rm max}$ is indeed smaller than $x_2^*$ by a term of 
order $O(\bar{g})$, as claimed above.
Obviously, since $a_1 < 1$, from Eq.\ (\ref{x2x1})
we derive the inequality $x_1^* < x_{\rm max} < x_2^*$, i.e.,
the gap function assumes its maximum between the values
$x_1^*$ and $x_2^*$. The value of the gap function at $x_{\rm max}$
can be estimated via a calculation similar to the one for the
difference $\phi_{0,2} - \phi_{0,1}$ above. The result is
$\phi_{max} \simeq \phi_0 \, [1+O(\bar{g}^2)]$. 
This means that the gap function
is fairly flat over a region of size $O(\bar{g})$ (in the variable
$x$) in the vicinity of the Fermi surface.

\section{Transition Temperature} \label{transtemp}

In this section we compute the transition temperature
$T_c$ where the color-superconducting condensate melts.
In the 2SC phase, in the CSL phase with longitudinal
gap, and in the polar phase, the calculation
of Ref.\ \cite{wang} applies, and we obtain
the BCS result $T_c/\phi_0 = e^\gamma/\pi$.
In these cases, $(\lambda_1^{a_1} \lambda_2^{a_2})^{1/2}=1$, 
cf.\ Table \ref{table1}, such that Eq.\ (\ref{Tc}) is valid.

In the transverse CSL phase, we also obtain the
BCS result for the relationship between $T_c$ and
the zero-temperature gap after
a rescaling of the gap function by a factor $\sqrt{\lambda_1}$,
cf.\ the discussion in Sec.\ \ref{solution}.
Since in this case $(\lambda_1^{a_1} \lambda_2^{a_2})^{1/2} =
\sqrt{\lambda_1}$, Eq.\ (\ref{Tc}) also applies.

In the CFL phase and the CSL phase with both longitudinal and
transverse gaps, we have
to compute $T_c$ explicitly. The calculation follows the line
of arguments presented in Ref.\ \cite{wang}, taking into account
the additional term $\sim a_1$ in Eq.\ (\ref{phix}).
As in Refs.\ \cite{rischke1,wang} we assume that, to leading
order, the effect of temperature is a change of the magnitude
of the gap, but not of the shape of the gap function,
\be
\phi(x,T) \simeq \phi(T)\, \frac{\phi(x,0)}{\phi_0} \;,
\ee
where $\phi(T)\equiv\phi(x_2^*,T)$ is the value of the gap 
at the Fermi surface at temperature $T$,
$\phi(x,0)$ is the zero-temperature gap function $\phi(x)$ computed in the
last section, cf.\ Eq.\ (\ref{phiz}),
and $\phi_0 \equiv \phi_{0,2} = \phi(x_2^*,0)$. 
With this assumption, Eq.\ (\ref{phix}) reads at the Fermi surface
\begin{eqnarray}
1 & = & \int_{x_0}^{x_\k} dy\, y\, (1- 2\, \bar{g} \, y) \,
\tanh\left[\frac{\e(y)}{2T}\right]  \, \frac{\phi(y,0)}{\phi_0}
+ \int_{x_\k}^{x_2^*}dy\, y\, (1- 2\, \bar{g} \, y) \,
\tanh\left[\frac{\e(y)}{2T}\right]  \, \frac{\phi(y,0)}{\phi_0}
\nonumber \\
&   & - a_1\, x_2^*  \int_{x_1^*}^{x_2^*}dy\, (1- 2\, \bar{g} \, y) \,
\tanh \left[ \frac{\e(y)}{2T} \right]\, \frac{\phi(y,0)}{\phi_0}
\nonumber \\
& \equiv & {\cal I}_1 + {\cal I}_2 + {\cal I}_3
\,\, ,
\label{gapeqtemp}
\end{eqnarray}
where we divided the second integral in Eq.\ (\ref{phix}) 
into two integrals: ${\cal I}_1$ which runs from 
$x_0$ to $x_\kappa$, with $x_\kappa \equiv
x_2^*-\bar{g}\, \ln(2\kappa)$, $\kappa \gg 1$, and ${\cal I}_2$ which runs 
from $x_\kappa$ to $x_2^*$ \cite{rischke1}. 
We now compute the integrals ${\cal I}_1$ through ${\cal I}_3$
separately to subleading accuracy, i.e., to order $O(\bar{g})$.

In the first integral ${\cal I}_1$, which runs over a region far from the
Fermi surface, $\e(y) \gg T$, and we may approximate the $\tanh$ by 1.
This integral can be formally solved by integration by parts using the
differential equation (\ref{diffeq}),
\be
{\cal I}_1 = \frac{1}{\phi_0} \, \left[ \phi(x_\k,0) - x_\k \,
\frac{d\phi}{dx}(x_\k,0) \right] \,\,,
\ee
where we exploited the condition (\ref{relation3}).
Expanding the functions on the right-hand side around $x_2^*$ we
obtain to subleading order
\be \label{I1}
{\cal I}_1 = 1 - \frac{\pi}{2} \,\left[ \bar{g}\, \ln (2 \k) - a_1 \,
(x_2^*-x_1^*) \right]\,\, .
\ee
This estimate is similar to the one made in Eq.\ (36) of Ref.\
\cite{wang}. The main difference to that calculation is the term
$\sim a_1$ which appears because the first derivative of the
gap function no longer vanishes at the Fermi surface, cf.\ the
discussion in the previous section.

In the second integral ${\cal I}_2$, 
which only contributes to order $O(\bar{g})$
to the right-hand side of Eq.\ (\ref{gapeqtemp}), to subleading order
we may set $\phi(y,0)/\phi_0 \simeq 1$ and $y \simeq  x_2^* \simeq
\pi/2$.
Reverting the transformation of variables (\ref{vartrans}) 
we obtain
\be
{\cal I}_2 = \frac{\pi}{2} \bar{g} \int_0^{\sqrt{\lambda_2}\k\phi_0} 
\frac{d(q-\m)}{\e_{q,2}}\, \tanh  \left( \frac{\e_{q,2}}{2T} \right) \,\, .
\ee

The last integral in Eq.\ (\ref{gapeqtemp}), ${\cal I}_3$, 
also contributes a term of order $O(\bar{g})$, 
and may thus be approximated by an argument
similar to that leading to Eq.\ (\ref{subcorr}),
\be
{\cal I}_3 = a_1\, x_2^* (x_2^* - x_1^*) \, 
\tanh \left[ \frac{\phi(T)}{2T} \right] \,\, .
\ee
At the critical temperature $T_c$, where $\phi(T_c) = 0$, 
this term vanishes. 
Putting everything together, at $T=T_c$ Eq.\ (\ref{gapeqtemp}) becomes
\be \label{gapeqtemp2}
\bar{g} \int_0^{\sqrt{\lambda_2}\k\phi_0} 
d(q-\m)\, \left[ \frac{1}{q-\m}\, \tanh 
\left( \frac{q-\m}{2T_c} \right) - 
\frac{1}{\sqrt{(q-\m)^2 + \lambda_2 \phi_0^2}} \right] = 
- a_1 \, (x_2^*- x_1^*) \,\, ,
\ee
where the term $\ln (2\k)$ in Eq.\ (\ref{I1}) was expressed
in terms of an integral according to Eq.\ (96) of Ref.\ \cite{rischke1}.
In the integral on the left-hand side, we may send $\k \rightarrow
\infty$ \cite{rischke1}. This allows us to perform it analytically,
which yields the result 
$\ln[ e^\gamma \sqrt{\lambda_2}\phi_0/(\pi T_c)]$, 
where $\gamma \simeq 0.577$ is the Euler-Mascheroni constant.
If the right-hand side of Eq.\ (\ref{gapeqtemp2}) 
were zero, for $\lambda_2 = 1$
this would then lead to the BCS relation $T_c/\phi_0 = e^\gamma/\p$.
However, using Eq.\ (\ref{x2x1}) we now obtain Eq.\ (\ref{Tc}).
The last factor on the right-hand side of this equation
is exactly the inverse of the additional 
factor in Eq.\ (\ref{phi0}). 
This factor violates the BCS relation $T_c/\phi_0 = e^\gamma/\pi$
in the CFL and CSL cases. In the first case, the transition temperature
is by a factor $2^{1/3}$ larger than one would expect from 
BCS theory and in the second case it is larger by a factor $2^{2/3}$. 
However, in units of energy,
this factor just cancels the one from $\phi_0$ in Eq.\ (\ref{phi0}).

\section*{Acknowledgments}

The authors thank P.\ Kopietz, M.\ Lang, P.\ Reuter, and 
T.\ Sch\"afer for interesting discussions. Q.W.\ acknowledges
support from the Alexander von Humboldt-Foundation.
This work was supported by BMBF and GSI Darmstadt.

\begin{appendix}

\section{Computing eigenvalues} \label{AppA}

The eigenvalues $\lambda_r$ of $L_{\bf k}$ follow from the roots of 
\be
{\rm det} \left( \lambda\, {\bf 1} - L_{\bf k}  \right) = 0 \,\, .
\ee
The left-hand side of this equation can be rewritten in the form
\be
{\rm det} \left( \lambda\, {\bf 1}  - L_{\bf k} \right) \equiv \exp
\left\{ {\rm Tr} \left[ \ln \left( \lambda\, {\bf 1} - L_{\bf k}  \right) 
\right]\right\} \,\, .
\ee
The logarithm of the matrix $\lambda\, {\bf 1} - L_{\bf k}$ is formally
defined in terms of a power series,
\be
{\rm Tr} \left[ 
\ln \left( \lambda\, {\bf 1} - L_{\bf k}  \right) \right] = \ln \lambda 
\, {\rm Tr}\, {\bf 1}
+ {\rm Tr} \left[ \ln \left( 1 - \frac{L_{\bf k}}{\lambda} \right) \right]
= \ln \lambda \, {\rm Tr}\, {\bf 1} 
- \sum_{n=1}^{\infty} \frac{1}{n}\, \lambda^{-n} \,
{\rm Tr}\, L_{\bf k}^n \,\, .
\ee
In order to proceed, one needs to know the trace of the $n$th power of the
matrix $L_{\bf k}$.
In the 2SC phase, the CSL phase with longitudinal gap, and in
the polar phase, this is particularly simple, since
$L_{\bf k}$ is a projector, cf.\ Eqs.\ (\ref{L2SC}),
(\ref{LCSLlong}), and (\ref{Lpolar}), hence
$L_{\bf k}^n \equiv L_{\bf k}$. Counting color and flavor degrees
of freedom in the 2SC phase, and color and Dirac degrees of
freedom in the longitudinal CSL and polar phases, the trace of $L_{\bf k}$ is
4 in the former and 8 in the latter case, respectively.
Therefore, we obtain for the 2SC phase
\be 
\mbox{\rm 2SC phase:} \qquad
{\rm det} \left( \lambda\, {\bf 1} - L_{\bf k} \right) = 
\lambda^2 \, (\lambda - 1)^4 =0\,\, .
\ee
This yields the eigenvalues given in Eq.\ (\ref{EV2SC}).
For the longitudinal CSL and polar phases we analogously compute
\be 
\mbox{\rm long.\ CSL and polar phases:} \qquad
{\rm det} \left( \lambda\, {\bf 1} - L_{\bf k} \right) = 
\lambda^4 \, (\lambda - 1)^8 =0\,\, ,
\ee
which leads to the eigenvalues of Eqs.\ (\ref{EVlong}) and
(\ref{EVpolar}).

The next simple case is the CSL phase with transverse gap.
In this case, $L_{\bf k}$ is not a projector, but 
since $\gperp \cdot \gperp = -2$, it is still
idempotent up to a factor, $L_{\bf k}^2 = 2 \, L_{\bf k}$.
Because of $L_{\bf k}^n = 2^{n-1} \, L_{\bf k}$ and ${\rm Tr}\, 
L_{\bf k} = 16$ we then obtain
\be 
\mbox{\rm trans.\ CSL phase:} \qquad
{\rm det} \left( \lambda\, {\bf 1} - L_{\bf k} \right) = 
\lambda^4 \, (\lambda - 2)^8 =0\,\, ,
\ee
from which we read off the eigenvalues of Eq.\ (\ref{EVtrans}).

The CFL phase and the CSL phase (with both longitudinal and transverse gaps)
are the only cases where the calculation of $L_{\bf k}^n$ is slightly
more involved. First, one proves the identity $L_{\bf k}^2 = 5\,
L_{\bf k} - 4 \, {\bf 1}$, which is valid in both cases.
Repeated application of this relation allows to reduce an arbitrary
number of powers of $L_{\bf k}$ to a single power, plus a term 
proportional to the unit matrix,
\be
L_{\bf k}^n = a_n\, L_{\bf k} + b_n \,{\bf 1} \,\, .
\ee
Multiplying both sides of this equation by $L_{\bf k}$,
one derives the recursion relation
\be \label{recurs}
a_{n+1} = 5\, a_n - 4\, a_{n-1}
\ee
for the coefficients $a_n$, and the identity
\be
b_{n+1} = - 4\, a_n
\ee
for the coefficients $b_n$. The recursion relation (\ref{recurs})
can be solved with the Ansatz $a_n = p^n$, which yields a quadratic
equation for $p$ with the solutions $p_1 = 4$ and $p_2 = 1$. 
The general solution of the recursion relation is
then $a_n = \alpha \, p_1^n + \beta\, p_2^n = \alpha\, 4^n + \beta$.
The coefficients $\alpha$ and $\beta$ can be determined from $a_1=1$ 
and $a_2=5$, such that
\be 
a_n = \frac{4^n - 1}{3} \quad, \qquad
b_n = - \frac{4^n - 4}{3} \,\, .
\ee
In the CFL phase, ${\rm Tr} \, L_{\bf k} = 12$ and
${\rm Tr}\, {\bf 1} = 9$, while in the
CSL phase, ${\rm Tr} \, L_{\bf k} = 24$ and 
${\rm Tr} \, {\bf 1} = 12$. Consequently, in the CFL phase
\be 
\mbox{\rm CFL phase:} \qquad
{\rm det} \left( \lambda\, {\bf 1} - L_{\bf k} \right) = 
(\lambda-4) \, (\lambda - 1)^8 =0\,\, ,
\ee
which leads to Eq.\ (\ref{EVCFL}), while in the CSL phase
\be 
\mbox{\rm CSL phase:} \qquad
{\rm det} \left( \lambda\, {\bf 1} - L_{\bf k} \right) = 
(\lambda-4)^4 \, (\lambda - 1)^8 =0\,\, ,
\ee
which yields Eq.\ (\ref{EVCSL}).

\section{Integration over gluon momentum} \label{AppB}

In this appendix we compute the integrals over
gluon 3-momentum $p$ to subleading order in the gap equation.
We shall see that to this order it is consistent to put
$k = q = \m$.

After replacing $\uk \cdot \uq = (k^2+q^2 -p^2)/(2kq)$,
the coefficients $\eta_{2m}^{\ell, t}(ee',k,q)$
can be read off from Eqs.\ (\ref{T2SC2}), 
(\ref{TCFL}), (\ref{TCSL}), (\ref{TCSLlong}), (\ref{TCSLtrans}),
and (\ref{Tpolar}). One first observes that for all cases considered
here, $\eta_{2m}^{\ell, t}(ee',k,q) = 0$ for $m \geq 3$.
Next, one also realizes that $\eta_{-2}^\ell = 0$, since
there is no term in ${\cal T}_{00}^{ee', i}$ proportional
to $1/p^2$. Consequently, we have to compute the integrals
\begin{mathletters}
\be \label{statelec}
{\cal I}_{2m}^\ell = \int_{|k-q|}^{k+q} dp \, p \,
\frac{2}{p^2+3m_g^2} \, \left( \frac{p^2}{kq} \right)^m \,\, , \;\;\;
m = 0,1,2\,\, ,
\ee
for the contribution of static electric gluons to the gap equation,
\be \label{nonstatmagnet}
{\cal I}_{2m}^{t,1} = \int_{M}^{k+q} dp \, p \,
\frac{2}{p^2} \, \left( \frac{p^2}{kq} \right)^m \,\, ,
\;\;\; m = -1,0,1,2\,\, ,
\ee
for the contribution of non-static magnetic gluons, and 
\be \label{statmagn}
{\cal I}_{2m}^{t,2} = \int_{|k-q|}^{M} dp \, p \,
\frac{p^4}{p^6 + M^4 \omega_\pm^2} \, \left( \frac{p^2}{kq} \right)^m \,\, ,
\;\;\; m = -1,0,1,2\,\, ,
\ee
\end{mathletters}
with $\omega_\pm \equiv \epsilon_{q,s}^{e'} \pm \epsilon_{k,r}^e$,
for the contribution of almost static magnetic gluons.
The result for the integrals (\ref{statelec}) and (\ref{nonstatmagnet}) is
\begin{mathletters} \label{results1}
\bea
{\cal I}_0^\ell & = & 
\ln \left[ \frac{(k+q)^2 + 3 m_g^2}{(k-q)^2 + 3 m_g^2} \right]
\simeq \ln \left( \frac{4\m^2}{3m_g^2} \right) \,\, , \\
{\cal I}_2^\ell & = & 4 - \frac{3 m_g^2}{kq}\,
\ln \left[ \frac{(k+q)^2 + 3 m_g^2}{(k-q)^2 + 3 m_g^2} \right] 
\simeq 4\,\, , \\
{\cal I}_4^\ell & = & 4 \, \frac{k^2 + q^2 - 3m_g^2}{kq}
+ \left( \frac{3 m_g^2}{kq} \right)^2\,
\ln \left[ \frac{(k+q)^2 + 3 m_g^2}{(k-q)^2 + 3 m_g^2} \right] 
\simeq 8\,\, , \\
{\cal I}_{-2}^{t,1} & = & \frac{kq}{M^2} - \frac{kq}{(k+q)^2} 
\simeq \frac{\m^2}{M^2} - \frac{1}{4}\,\, , \label{I-2t1}\\
{\cal I}_0^{t,1} & = & \ln \left[ \frac{(k+q)^2}{M^2} \right] 
\simeq \ln \left( \frac{4 \m^2}{M^2} \right)\,\, , \\
{\cal I}_2^{t,1} & = & \frac{(k+q)^2-M^2}{kq}
\simeq 4\,\, , \\
{\cal I}_4^{t,1} & = &  \frac{(k + q)^4 - M^4}{2(kq)^2} 
\simeq 8\,\, . 
\eea
\end{mathletters}
The approximate equalities on the right-hand sides hold
to subleading order in the gap equation. One obtains them
employing two approximations. First, terms proportional to
at least one power of $m_g^2$ or $M^2$ carry at least two
additional powers of $g$, 
which renders them sub-subleading and thus negligible
to the order we are computing. Second, one
utilizes the fact that the $q$ integration in the gap equation
is over a region of size $2 \delta$ around the Fermi surface,
where $\delta \sim m_g$. To subleading order it is thus
accurate to put $k = q = \mu$ (see discussion in Sec.\
\ref{generalder}). This then yields the right-hand sides
of Eqs.\ (\ref{results1}).

Note that there is a term $\sim \m^2/M^2 \sim 1/g^2$ in Eq.\
(\ref{I-2t1}). This term is parametrically the
largest and could in principle give the dominant contribution to the
gap equation. However, in all cases considered here, it turns out that
the coefficient $\eta_{-2}^t$ is proportional to at least one power
of $(k-q)^2$. Performing also the $q$ integration in the
gap equation, one then has terms of the form
\be \label{estimate}
g^2 \int_0^\delta \frac{d(q-\m)}{\epsilon_q}\, \frac{(k-q)^2}{M^2} \, 
\phi (\epsilon_q,q)
\sim g^2\, \frac{\phi_0}{M^2} 
\int_0^\d \frac{d \xi}{ \sqrt{\xi^2 + \phi_0^2}}\, \xi^2
\sim g^2 \, \phi_0 \, \frac{\d^2}{M^2} + O\left(\frac{\phi_0^3}{\m^2}
\right)\,\, ,
\ee
where for the purpose of power counting we have neglected
the $q$ dependence of the gap function, $\phi(\e_q,q) \sim \phi_0$,
and we have evaluated the integral on the left-hand side for
$k = \m$.
As long as $\d \sim m_g \sim M$, the leading term
in Eq.\ (\ref{estimate}) is $\sim g^2 \phi_0$,
and thus it is only of sub-subleading order in the gap equation. 
It is obvious that the constant term $- 1/4$ in Eq.\ (\ref{I-2t1})
is parametrically even smaller. The contribution to the term
$\sim \eta_{-2}^t$ from non-static magnetic gluons is therefore
negligible to subleading order. 

Finally, also the integrals ${\cal I}_{2m}^{t,2}$ can be
computed analytically \cite{gradstein}. Defining
$\a \equiv (M^4 \omega_\pm^2)^{1/3}$, the result is
\begin{mathletters}
\bea
{\cal I}_{-2}^{t,2} & = & - \frac{kq}{12\a} \,  \left\{ 
\ln \left[\frac{(x+\a)^2}{x^2- \a x + \a^2}\right] - 
2 \sqrt{3}\, {\rm arctg}\, \left( \frac{2x-\a}{\sqrt{3}\,\a} \right) 
\right\}^{M^2}_{(k-q)^2}  \,\, , \label{B1} \\
{\cal I}_0^{t,2} & = & \frac{1}{6} \, 
\ln \left[ \frac{M^6 + \a^3}{(k-q)^6 + \a^3}
\right] \simeq \frac{1}{6} \, \ln \left(
\frac{M^2}{\omega_\pm^2}\right)\,\, , \label{B2} \\
{\cal I}_2^{t,2} & = & \frac{M^2 - (k-q)^2}{2kq} - \frac{\a}{12kq}\, 
\left\{ 
\ln \left[\frac{(x+\a)^2}{x^2- \a x + \a^2}\right] + 2 \sqrt{3}\, {\rm arctg}\,
\left( \frac{2x-\a}{\sqrt{3}\,\a} \right) \right\}^{M^2}_{(k-q)^2} 
\simeq 0 \,\, ,  \label{B3} \\
{\cal I}_4^{t,2} & = & \frac{M^4-(k-q)^4}{4(kq)^2} -
\frac{\a^3}{(kq)^3} 
\, {\cal I}_{-2}^{t,2}
\simeq 0 \,\, . 
\label{B4}
\eea
\end{mathletters}
Here, we used the short notation $\{ f(x) \}^{a}_{b} \equiv
f(a) - f(b)$. In order to obtain the approximate equalities on 
the right-hand sides of Eqs.\ (\ref{B2}), (\ref{B3}), and (\ref{B4}),
one employs the fact that typically $(k-q)^2 \sim \omega_\pm^2 \ll M^2$, 
such that parametrically $(k-q)^2 \ll \a \ll M^2$. 
This immediately yields the right-hand side of Eq.\ (\ref{B2}).
For Eqs.\ (\ref{B3}) and (\ref{B4}), we use this estimate in order to
expand the logarithm occurring in Eqs.\ (\ref{B1}) and
(\ref{B3}). One finds that the leading term is $\sim \a/M^2$.
Similarly, one expands the inverse tangent occurring in these equations,
which leads to terms which are even of order $O(1)$. 
Collecting all prefactors, however, all terms in Eqs.\
(\ref{B3}) and (\ref{B4}) are then suppressed by at least one
power of $g^2$. These sub-subleading corrections are negligible
to the order we are computing.

Somewhat more care is necessary in estimating the terms in Eq.\
(\ref{B1}).
Again, one may expand the logarithm and the inverse tangent.
Together with the prefactor, this leads to a term $\sim 1/M^2$ for
the logarithm, and a term $\sim 1/\a$ for the inverse tangent.
The first term is harmless: together with the factor $(k-q)^2$ from
$\eta_{-2}^t$ it leads to an integral of the form (\ref{estimate}),
which was already shown to give a sub-subleading contribution to
the gap equation. The other term leads to the integral
\be
g^2 \int_0^\d \frac{d (q-\m)}{\e_q} \, 
\frac{ (k-q)^2}{\a} \, \phi(\epsilon_q,q)
\sim g^2 \, \frac{\phi_0 }{M^{4/3}} \int_0^\d 
\frac{d\x\, \x^2}{(\x^2 + \phi_0^2)^{5/6}}\,\, ,
\ee
where we used similar power-counting arguments as in Eq.\
(\ref{estimate}).
The last integral is finite even for $\phi_0=0$, so
that we can estimate it to be $\sim \delta^{4/3}$.
For $\d \sim m_g$ this contribution is then again $\sim g^2 \phi_0$
and thus of sub-subleading order in the gap equation.

In conclusion, also the contribution of
almost static magnetic gluons to the term $\sim \eta_{-2}^t$
is of sub-subleading order and can be neglected.
To subleading order, it is therefore consistent to put $\eta_{-2}^t =
0$ from the beginning, provided one chooses $\d \sim m_g$.

\section{Dirac structure of the spin-1 gap matrix} \label{AppD}

Since the gap matrix $\Sigma_{21}(K)$ is a 
complex $4\times 4$ matrix in Dirac space, 
it can be written as a linear combination of
sixteen basis matrices. The gap matrix is a scalar in momentum space, 
which reduces this number to eight \cite{rischke2}. 
We choose this basis set to be
\be
{\bf M}\equiv (M_1,\ldots,M_8)\equiv({\bf 1},\gamma^0,
\vg\cdot\uk,\vg\cdot\uk\,\gamma^0,\vg\cdot\uk\,\gamma^0\gamma^5,
\gamma^0\gamma^5,\vg\cdot\uk\,\gamma^5,\gamma^5) \;.
\ee
In the spin-1 case the order parameter 
has to be a three-dimensional vector.
But $\Sigma_{21}$ has no vector structure. Thus
the order parameter has to be contracted with other vectors. 
The only available vectors are $\uk$ and $\vg$, and each
contraction can still multiply any element of ${\bf M}$.
Thus, we can write the gap matrix in terms of 16 3-vector
order parameters $\vphi_i$, $i=1, \ldots, 16$, as 
\be
\Sigma_{21}(K)=\sum_{i=1}^{8}\vphi_i(K)\cdot\uk \, M_i 
+\sum_{i=9}^{16}\vphi_i(K)\cdot\vg\, M_{i-8} \; .
\ee
Decomposing the $i$th vector order parameter $\vphi_i$
into a longitudinal and a transverse part with respect to $\uk$, 
$\vphi_i=\varphi_i^\ell\, \uk+\vphi_i^t$, 
where $\vphi_i^t=\vphi_i\cdot(1-\uk\,\uk)$, the gap matrix 
becomes 
\be
\Sigma_{21}(K)=\sum_{i=1}^{8}\varphi_i^\ell(K) \, M_i 
+\sum_{i=9}^{16} \left[ \varphi_i^\ell (K)\, \vg
\cdot\uk+\vphi_i^t(K)\cdot\vg \right]\, M_{i-8} \;.
\ee
Note that whenever a basis matrix $M_i$ is multiplied (from the left)
by $\vg\cdot\uk$ one obtains, up to a minus sign, another basis matrix $M_j$.
Thus we can rearrange this expression such that there are
only eight different longitudinal coefficients. 
The eight longitudinal and sixteen (independent) transverse coefficients
can be combined into eight new 3-vector
order parameters $\vf_i$, 
\be
\Sigma_{21}(K)=\sum_{i=1}^8\vf_i(K) \cdot 
\left[ \uk + \gperp({\bf k}) \right]\, M_i\;,
\ee
where $\gperp({\bf k}) \equiv \vg \cdot (1-\uk\, \uk)$.
Now one can perform a basis transformation and 
write the gap matrix in terms of the projectors 
for energy, chirality and helicity. 
This is completely analogous to the spin-0 case \cite{rischke2}. 
In the ultrarelativistic limit two out of the
three projectors are sufficient, for instance those for energy, 
$\Lambda_{\bf k}^e$, and chirality, ${\cal P}_h=(1+h \gamma_5)/2$,
$h=\pm$ for right- or left-handed quarks, respectively.
Then we are left with four 3-vector order parameters
in terms of which the gap matrix reads
\be
\Sigma_{21}(K)=\sum_{e,h}\vf_h^e(K)\cdot \left[ \uk+\gperp({\bf k})\right]
\,{\cal P}_h\, \Lambda^e_{\bf k}\;.
\ee
For condensation in the even-parity channel, only two 3-vector
order parameters are independent, because
$\vf_r^e = \vf_\ell^e \equiv \vf^e$
\cite{rischke1}. The sum over chirality projections can
be immediately performed to give
\be
\Sigma_{21}(K)=\sum_{e}\vf^e(K)\cdot \left[ \uk+\gperp({\bf k}) \right]\,
\Lambda^e_{\bf k}\;.
\ee
This is Eq.\ (\ref{gmspin1}).

\end{appendix}

\end{document}